# THE COLLECTED PAPERS OF

# Albert Einstein

## VOLUME 15

### THE BERLIN YEARS:
### WRITINGS & CORRESPONDENCE
### JUNE 1925–MAY 1927

Diana Kormos Buchwald, József Illy, A. J. Kox,
Dennis Lehmkuhl, Ze'ev Rosenkranz, and Jennifer Nollar James
EDITORS
Anthony Duncan, Marco Giovanelli, Michel Janssen,
Daniel J. Kennefick, and Issachar Unna
ASSOCIATE & CONTRIBUTING EDITORS
Emily de Araújo, Rudy Hirschmann, Nurit Lifshitz, and Barbara Wolff
ASSISTANT EDITORS

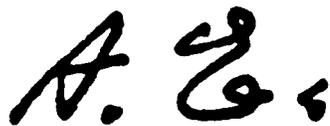

Princeton University Press
2018

LIBRARY OF CONGRESS CATALOGING-IN-PUBLICATION DATA
(*Revised for volume 15*)

Einstein, Albert, 1879–1955.
The collected papers of Albert Einstein.

German, English, and French.
Includes bibliographies and indexes.
Contents: v. 1. The early years, 1879–1902 / John Stachel, editor —
v. 2. The Swiss years, writings, 1900–1909 — — v. 15. The Berlin years,
writings and correspondence, June 1925–May 1927 / Diana Kormos Buchwald… [et al.], editors.
QC16.E5A2    1987      530      86-43132
ISBN 0-691-08407-6 (v.1)
ISBN 978-0-691-17881-3 (v. 15)
This book has been composed in Times.

The publisher would like to acknowledge the editors of this volume for
providing the camera-ready copy from which this book was printed.

Princeton University Press books are printed on acid-free paper
and meet the guidelines for permanence and durability of the
Committee on Production Guidelines for Book Longevity
of the Council on Library Resources.

Printed in the United States of America

1  3  5  7  9  10  8  6  4  2

The present volume covers a thrilling two-year period in twentieth-century physics, for during this time matrix mechanics—developed by Werner Heisenberg, Max Born, and Pascual Jordan—and wave mechanics, developed by Erwin Schrödinger, supplanted the earlier quantum theory. In extensive exchanges with the creators of the new approaches, Einstein quickly recognized their great importance and the conceptual peculiarities involved. From the beginning he preferred wave mechanics over matrix mechanics. He thought he had found a convincing refutation of the probabilistic interpretation of quantum mechanics in what would today be called a hidden variable theory, but he retracted the paper before publication.

Einstein also continued to work on other topics. In early 1925 he had turned to a new mathematical foundation of unified field theory that generalized Arthur S. Eddington's affine approach on which most of his previous attempts at a unified theory had been based. But he soon abandoned this approach, and in 1927 returned to a different one that he had earlier dismissed: the idea of Theodor Kaluza, further developed by Oskar Klein, that gravity and electromagnetism can be unified by introducing a fifth spacetime dimension. Between these two approaches, and inspired by detailed correspondence with the mathematician G. Y. Rainich, Einstein explored features of general relativity in the hope of finding new hints at how the correct unified field theory might look. This correspondence eventually brought about the important Einstein-Grommer paper of 1927, in which they aimed to derive the motion of particles subject to gravitational fields from the gravitational field equations themselves.

At the same time, Einstein discussed the interpretation of general relativity and unified field theories with the philosopher Hans Reichenbach. It is here that we find the first statements expressing his decade-long opposition to the idea that general relativity shows that gravity is "geometrized."

In a collaboration with Emil Rupp, Einstein became convinced that Rupp's experiments showed that excited atoms emitted light in a finite time (in waves) rather than instantly (in quanta). However, in subsequent years Rupp's experiments could not be reproduced.

Much of Einstein's correspondence in this volume engages with Dayton C. Miller's interferometric experiments in which he claimed to have detected an ether drift, overturning the null result of the Michelson-Morley experiment and generating renewed interest in experiments of this type in both Europe and the United States.

As in the past, relativity remained a contested topic among right-wing circles in Germany and abroad.[1] In March 1927, Einstein learned that a high school teacher in Virginia had been charged with blasphemy for teaching relativity. In his sarcastic retort, Einstein lampooned the school's directors, pointing out that they were so lacking in confidence that they needed God's help to assist them in their campaign against relativity (Doc. 493).

The current volume encompasses a wealth of documents, ranging over several significant scientific topics, as well as politics, Zionism, and myriad family concerns. We present 535 documents as full text and more than 900 documents in the Calendar of Abstracts. Among the former are 99 writings, of which only 56 have previously been published. They include two dozen scientific papers, drafts, and calculations, as well as poems, aphorisms, homages to Isaac Newton and Hendrik A. Lorentz, more than three dozen appeals and writings on political matters and Jewish affairs, and several patents. Among the 440 letters presented as full text, 270 were written by Einstein. This massive personal and professional correspondence of more than 1,300 letters, and the almost 100 writings show that Einstein's immense productivity and hectic pace of life were more intense during the twenty-four months covered by this volume than in the previous two years.[2]

He undertook several unsuccessful attempts to reduce his involvement in various spheres of activity and to balance private life, work, and public roles. In mid-June 1925, Einstein informed Mileva Marić that he felt well after his South American trip because the return voyage had been "so restful" (Doc. 7). However, merely eight days later, he wrote to Paul Ehrenfest and others that he did not intend to travel either to Pasadena or to Petrograd, as he needed to be "more frugal with his nerves" (Docs. 10, 20, and Abs. 95). During 1926, Einstein attempted to lighten the burden of responsibilities. In January, he offered his resignation from the board of the German League of Human Rights (Doc. 149), but eventually decided to remain on it. He also informed the Marxist-Zionist party Poale Zion that he would no longer support multiple individual Jewish causes since the overuse of his name would lead to its devaluation (Doc. 150). In this spirit, he also let the World Union of Jewish Students know that he had "resigned [his] honorary position as king of the schnorrers for good" (Doc. 196).

The 1925 Locarno Treaties renewed Einstein's optimism in the prospects for European reconciliation. He continued to participate in the League of Nations' International Committee on Intellectual Cooperation and efforts to end the boycott of German scientists. He also remained committed to the shaping of the Hebrew University in Jerusalem, although his enthusiasm for this cause was sorely tested during these years.

Einstein received many honors, among them the Royal Society's Copley Medal (Doc. 102, see Illustration 23), the Royal Astronomical Society's gold medal (Doc. 178), and election as corresponding member of the Academy of Sciences of the USSR (Abs. 668). He was also offered a faculty position at Johns Hopkins University (Doc. 465).

## I.  Romance with Marie Winteler

This volume contains hitherto unknown, much earlier correspondence between the sixteen-year-old Einstein and members of the Winteler family, with whom he lodged while attending the Aargau Kantonsschule in 1895–1896. In 2015, the Bernisches Historisches Museum made accessible a bundle of letters and postcards written by Einstein that had been obtained from the Winteler family estate.[3] Most of these letters are addressed to the eighteen-year-old Marie Winteler, his hosts' daughter, with whom he became romantically involved at the time (see Illustration 1). They also include a "Contract for the Purchase of a Box of Water Colors," drawn up with great, yet most likely, mock seriousness, by Einstein and his cousin Robert Koch (Vol. 1, 16c). Some of the items only exist as fragments or snippets, as many were torn and subsequently glued back together. Prior to the release of this new correspondence, only one letter by Einstein to Marie, and two letters by Marie to Einstein, were known to scholars.[4] Two additional letters provided hints of the end of their relationship in 1897.[5]

The twenty new letters and postcards from his youth, presented in full text, and the fourteen letters in the Calendar of Abstracts reveal Einstein's passionate, affectionate, and playful sentiments for Marie. His earliest letters from Aarau date from the beginning of 1896, after Marie had taken up a temporary teaching position in the nearby town of Niederlenz. Even though she apparently returned home at least once a week, when the two young people could meet, Einstein's letters express his deep longing for her during her brief absences and while he was on vacation with his family in Pavia. Unfortunately, Einstein did not preserve Marie's letters to him.

The letters document what appears to be Einstein's first romantic love, with all the attendant highs and lows of adolescent passion. He was at times brought to tears by Marie's notes to him. He alternately believed himself unworthy of her love, or described her as his "comforting angel," or felt at one with her (Vol. 1, 16d, 16f, and 16g). He was worried that she might be undernourished and embarked on a "project" to help her gain weight by sending her sausages. He encouraged her to take frequent walks and play the piano (Vol. 1, 16b and 16f). At times, he also attempted to make Marie jealous by mentioning other young women he had encountered (Vol. 1, 16b).

Einstein mentioned to Marie his difficulties in being a disciplined correspondent, an issue he would often return to in later years (Vol. 1, 15a). He allows glimpses into his career ambitions as well: on the eve of his departure from Aarau to take up his university studies at the ETH, he reported a conversation with the rector of the Kantonsschule during which he was told that he possessed the prerequisites for an academic career and was advised not to take up a position as a schoolteacher (Vol. 1, 27a). This move to Zurich in the fall of 1896 seems to have spelled the end of their romantic relationship. By next spring, remorseful for the heartache he had caused Marie, Einstein bemoaned his fate as a "mere schoolboy" who had nothing to offer her. The specifics of their estrangement are lacking, but it does appear that, while both of them struggled, Einstein blamed himself for the end of their romance (Vol. 1, 31a and 33a). He tried to reestablish contact with Marie two years later, but the outcome of this attempt remains unclear (Vol. 1, 53b).

In a surprising turn of events hitherto unknown to scholars, three letters and one postcard written by Einstein in 1909–1910 reveal that his love for Marie was rekindled at that time, more than a decade after their first relationship had ended. They apparently had a brief romantic encounter in 1909, by which time Einstein had already been married for over six years to Mileva Marić. But Marie seems to have ignored his subsequent advances, eliciting feelings of utmost anguish in Einstein. In his despair, he wrote in September 1909 that he felt "as if dead in this life filled with obligations, without love and without happiness," decrying his "failed love, failed life, that's how it always reverberates to me" (Vol. 5, 177a and 198a). By the summer of 1910, most likely after learning of Marie's engagement, Einstein wrote: "it seemed to me as if I were watching my grave being dug. The residual joy that still remained for me has been destroyed" (Vol. 5, 218a).

## II.  Electron Solutions, The Problem of Motion, and Metric-Affine Field Theory

Yuri Germanovich Rabinovich was a young mathematician working at the University of Odessa when he was arrested in 1922. He escaped that same year and, together with his wife, left Russia via Istanbul. He eventually arrived in the United States, changed his name to George Y. Rainich, and became a postdoctoral fellow at Johns Hopkins University (see Illustration 29).[6] His first contact with Einstein was facilitated by Jerome Alexander, a colloid chemist, who had commissioned Rainich to translate Einstein's paper on critical opalescence (*Einstein 1910d* [Vol. 3, Doc. 9]) into English. Rainich not only translated the paper but also added line-by-line comments in which he pointed out typographical errors, weaknesses in the argument, and ways to improve it. In his three letters to Einstein (Abs. 93, 114, and 214), Alexander asked him to address Rainich's comments, and whether any changes to the original paper should be made; no answer from Einstein is extant.

Soon afterward, Rainich wrote to Einstein directly (Doc. 96). Quoting from the opening remarks of Einstein's latest attempt at a unified field theory of gravity and electromagnetism (*Einstein 1925t*, Doc. 17), in which Einstein had stated that a convincing theory had yet to be found, Rainich enclosed a paper of his own from the year before (*Rainich 1925a*) in which he believed he had shown that there is a way to provide the sought-after unification within the framework of Einstein's original theory of general relativity. This time Einstein answered swiftly (Doc. 106). But rather than engaging with Rainich's paper, he elaborated on how much it disturbed him that in the context of the 1915 Einstein equations with an electromagnetic source term, the electromagnetic and the gravitational field enter as separate entities.

In response, Rainich summarized his results on how and to what extent one could unify gravity and electromagnetism within the framework of general relativity, i.e., within a pseudo-Riemannian manifold subject to a version of the Einstein equations. His account relied on a decomposition of the Riemann tensor into a symmetric and an antisymmetric tensor; he related the former to gravity and the cosmological constant, and the latter to electromagnetism (Doc. 126).

Instead of being captured by this possibility, Einstein latched on to Rainich's mathematical result of decomposing the Riemann tensor in this way, and to the fact that Rainich had managed to relate both the gravitational field and the cosmological constant to the symmetric part of the curvature tensor. This must have prompted Einstein to return to his own modified field equations of 1919 (*Einstein 1919a*, Vol. 7, Doc. 17) in which he had first suggested an alternative to his field equations of late November 1915 with the total energy-momentum tensor given by that of the electromagnetic field. The alternative consisted in replacing the Einstein tensor $R_{\mu\nu} - \frac{1}{2}g_{\mu\nu}R$ by the trace-free tensor $R_{\mu\nu} - \frac{1}{4}g_{\mu\nu}R$. This allowed him to obtain the cosmological constant as a constant of integration and to reconceptualize it as a constant negative pressure term. He had hoped to find solutions to the modified field equations that would be capable of representing elementary particles, with the cosmological constant thus reinterpreted playing a role in ensuring the stability of the particles. At the same time, Einstein explicitly did not think of the 1919 equations as a unification of the gravitational and electromagnetic fields, as expressed at the time in a letter to Kaluza of 29 May 1919 (Vol. 9, Doc. 48).

Einstein had already revisited these equations in *Einstein 1922r* (Vol. 13, Doc. 387), where he had argued that the vacuum equations of 1919 are equivalent to the vacuum equations with cosmological constant introduced in *Einstein 1917b* (Vol. 6, Doc. 43).[7] Now, with Rainich's results in hand, Einstein saw the opportunity to find an equally "mathematically natural interpretation" to the left-hand side of the 1919 equations (and thus also to that of the 1917 equations), as Gustav Herglotz had done for the original Einstein tensor of 1915 (*Herglotz 1916*). In *Einstein 1927a* (Doc. 158), Einstein described how Herglotz had related the Einstein tensor to the median curvature of three-dimensional slices through space-time; now Einstein himself argued that the trace-free tensor of 1919 vanishes if and only if Rainich's antisymmetric part of the Riemann curvature tensor vanishes.

In a letter to Michele Besso (Doc. 138) he praised the 1919 field equations as the "best of what we have today," and gave a similar endorsement in a letter to Eddington. But he also pointed out that the question remained whether the equations, and with them general relativity, "fail in the face of quantum phenomena" (Doc. 179).

The first part of Einstein's correspondence with Rainich may be surprising in that Einstein was much less enthusiastic than one might expect. After all, if true, Rainich's results would mean that the unified field theory that Einstein had been seeking for years was already at hand. But Einstein's reaction was consistent with his previous conduct of this search. Again and again, Einstein had abandoned approaches to a unified field theory not because he felt that they were lacking as unifiers of the gravitational and the electromagnetic field, but because they did not *also* solve "the quantum problem" by allowing for solutions that could be interpreted as electrons. Indeed, in *Einstein 1924d* (Vol. 14, Doc. 170), Einstein had spelled out a research program he had already implicitly followed for years: every satisfactory unified field theory of gravity and electromagnetism needed to have "at least the static spherically symmetric solution that describes the positive, or the negative, electron."

Einstein had then recently developed another candidate unified field theory. In *Einstein 1925t* (Doc. 17), which he described to Ehrenfest as a captivating paper (see Doc. 71), he both continued and departed from what he had called "the Weyl-Eddington approach" to unified field theories. The approach starts from Hermann Weyl's and Tullio Levi-Civita's realization that the affine connection can be defined independently of the metric, and from Eddington's idea of basing a unified field theory only on the affine connection that recovers the metric as a derivative concept. Einstein had followed this doctrine in a series of papers published between 1921 and 1925, and investigated different candidate field equations for the affine connection (see Vol. 14, Introduction, sec. I). In *Einstein 1925t* (Doc. 17), he departed from the tenet of introducing only the connection as the fundamental object and instead introduced both an affine connection and a metric tensor as fundamental and independent objects on which to base the theory.[8] As in his previous papers on affine field theory, Einstein finished by noting that it would now be necessary to investigate whether the theory allows for spherically symmetric solutions that could describe electrically charged particles. In a letter to Besso (Doc. 34), he likewise alluded to the critical question of whether the theory predicts the existence of quanta. For Einstein this meant, first and foremost, that solutions representing electrons have a uniquely determined electric charge and rest mass, and that no continuum of charge and rest mass exists.

Alas, as before, Einstein soon lost faith in his latest approach. In letters to Paul Ehrenfest (Doc. 71), Eddington (Doc. 91), and Lorentz (Doc. 94), he abruptly disowned the theory; writing this to Lorentz must have been particularly galling, as Lorentz had just sent a ten-page letter in which he discussed the theory in meticulous detail (Doc. 90).

Abandoning yet another stillborn unified field theory renewed internal doubts regarding the very basis of Einstein's earlier attempts at unification. Instead of immediately returning to the fray with a modified approach he now endeavored to give a general argument why all these attempts *had to* fail. In his contribution to the Festschrift honoring the fiftieth anniversary of Lorentz's doctorate (*Einstein 1925w*, [Doc. 92]), Einstein aimed to prove a meta-theorem about a whole class of theories that included, as special cases, Einstein-Maxwell theory, Eddington's theory, and his own affine field theories, as well as his newest approach based on a mixed metric-affine geometry. The theorem stated that any theory that represents the electromagnetic field by an antisymmetric second rank tensor, and that has a solution representing a particle with the charge and mass of an electron, must also have a solution corresponding to a particle with the same mass but opposite (i.e., positive) electric charge. And since Einstein assumed, as did every physicist at the time, that the only two fundamental particles were the electron and the proton (with the latter about two thousand times heavier than the former), he judged that all these theories were empirically inadequate. Thus again we see that Einstein's litmus test for any field theory was that it should predict the discrete (quantum) features of the known elementary particles.

Einstein's Lorentz Festschrift paper led Rainich to write yet again. He had penned a one-page response (*Rainich 1926b*), and sent it to Einstein on the same day that he sent it for publication to Adriaan Fokker, the editor of the journal in which Einstein's paper (together with all the others presented on the occasion of Lorentz's anniversary) were published. Rainich disputed Einstein's interpretation of his meta-theorem, which brought about the second, even more fruitful, phase of their exchanges during which they discussed the difference between linear and nonlinear field equations, and between Einstein's program of searching for solutions that would be capable of representing both the interior and the exterior of electrons and Rainich's program of deriving the properties of such particles from their exterior fields (Docs. 216, 245, 258, 293, 300).

In the course of the correspondence, Einstein underwent a remarkable change. On 18 April 1926 (Doc. 258) he wrote that the "cardinal question is of course whether one should think of electricity as continuous or made up of singularities," two options he had already offered in his 1922 Princeton lectures (Vol. 7, Doc. 71). There he had deemed the latter option a pseudosolution (*Scheinlösung*), favoring the former option instead, just as he did in his correspondence with Rainich. However, on 6 June 1926, he wrote: "[This] is the core question: a theory is sensible only if it allows a derivation of the equations of motion of particles without any extra assumptions. Whether the electrons are treated as singularities or not does not really matter *in principle*" (Doc. 300).[9]

This line of thought culminated in the longest scientific writing in the present volume: a paper cowritten with Jakob Grommer on the problem of motion, i.e., on how to derive the equations of motion of particles directly from the gravitational field equations. The paper, *Einstein and Grommer 1927* (Doc. 443), starts out by contrasting three possible approaches (*Betrachtungsweisen*) to the problem, two of which are serious contenders. The question is whether to start from the full Einstein equations, with the energy-momentum tensor of generic material systems as a source term, and to derive the equations of motion via the Bianchi identities; or whether to start instead from the vacuum Einstein equations and allow for particles to correspond to singularities in the metric field. Einstein and Grommer opt for the latter alternative and start by investigating a special class of solutions to the vacuum field equations, the Weyl class of static axialsymmetric solutions. Einstein had cited Weyl's and Levi-Civita's papers on such solutions during his correspondence with Rainich (Doc. 258), arguing that they contained a static two-body solution, exactly the kind of solution that Rainich had claimed would render general relativity empirically inadequate because it would correspond to two bodies exerting gravitational fields on one another, yet not move toward each other (*Rainich 1926b*). But Rainich had kept insisting that a static two-body solution need not exist in a nonlinear theory such as general relativity (Doc. 293). Einstein ended up agreeing with him (Doc. 300). He must have gone back to these papers by Weyl and Levi-Civita, following Rainich's insistence, and it is plausible that this brought about the change in what constituted "the core question."

For now, in his paper with Grommer, Einstein examines these solutions in detail, and uses them to argue that, according to general relativity, it is impossible for a particle to be subject to an external gravitational field and yet remain at rest. The argument depends on effectively distinguishing between acceptable and unacceptable singularities in the spacetime metric. Einstein and Grommer allowed for singularities to correspond to (or serve as placeholders for) material particles, but they did not allow singularities in regions of spacetime free of matter. Weyl's papers on axialsymmetric solutions show that a solution to the vacuum Einstein equations capable of representing a static two-body system would need to allow for a line singularity between the two bodies. Thus, contrary to what Einstein thought when he first wrote of this solution to Rainich, upon closer inspection the solution must have been unacceptable to him as a *physical* solution representing two bodies at rest and interacting gravitationally. Einstein and Grommer modified the solution Weyl had interpreted as representing a static two-body system so that they could interpret it as representing one body subject to an external gravitational field. But the same argument applies: the solution is unacceptable because, in addition to the (acceptable) singularity corresponding to a material particle, it involves an (unacceptable) singularity along the rotation axis in spacetime regions free of matter.

Einstein and Grommer thus effectively concluded that there is no physical solution corresponding to a particle at rest but subject to an external gravitational field. They used this result to argue that, in general, the motions of particles can be determined from the vacuum field equations. As they note themselves, this would make general relativity unlike any of the theories that had preceded it (if, as did Einstein and Grommer, one disregards the approach featuring the energy-momentum tensor): for the first time, it would not be necessary to postulate both field equations and equations of motion for particles subject to the field in question. They began their argument by reformulating the vacuum Einstein equations in terms of a surface integral over a three-dimensional hypersurface, and defining gravitational energy-momentum flow through the surface. They then picked a curve that was supposed to represent the path of a material particle, imposed the linear approximation that the metric deviates only slightly from Minkowski spacetime, and noted one of the main points made by Rainich: that not all solutions to the linearized field equations will correspond to solutions of the nonlinear vacuum Einstein equations that the linearized field equations approximate.

However, they then proposed a potential solution to this problem that is not found in Einstein's correspondence with Rainich: *if* a certain "equilibrium condition" for the gravitational energy-momentum flowing through the three-surface defined previously is fulfilled, *then* a solution to the linearized field equations *will* also solve the full non-linear field equations. Assuming this condition, Einstein and Grommer split the metric in the neighborhood of the curve of the material particle into an "inner metric" (owing to the particle itself) and an "outer metric" (owing to other gravitational sources or the lack thereof). Importantly, the outer metric did not contain any singularities, while the inner metric was taken to be singular. Finally, they calculated the surface integral that is equivalent to the vacuum field equations "around" the curve of the material particle and concluded that this curve is a geodesic of the outer metric. Einstein and Grommer thus concluded that the geodesic motion of particles subject only to gravity followed from the field equations.[10]

This constituted a significant change. In all of Einstein's previous publications on relativity he had been careful to stress that the field equations and the equation of motion of particles subject only to gravity—the geodesic equation—needed to be postulated independently. However, he must have asked himself early on whether this was really necessary. For already in the *Entwurf* theory of 1913, Einstein and Marcel Grossmann had shown that for the special case where all the matter in a given spacetime region is pressureless dust, the condition that the covariant divergence of the energy-momentum tensor vanishes implies the equations of motion of dust particles (see *Einstein and Grossmann 1913* [Vol. 4, Doc. 13] and the Zurich Notebook [partially published as Vol. 4, Doc. 10]). In a document that was likely a draft for the 1921 Princeton lectures, Einstein stated that the field equation "already contains the divergence equation and with it the laws of motion of material particles" (Vol. 7, Doc. 63). But no such statement is contained in the final Princeton lectures; as before, Einstein introduced the field equations and the equations of motion as separate assumptions.

The likely reason was that Einstein was unhappy with the role the energy-momentum tensor played in these approaches; he had emphasized again and again that the energy-momentum tensor was only a phenomenological representation of matter, to be regarded with caution. In this volume, the clearest case is found in a letter to Besso from 11 August 1926, where he wrote: "But it is questionable whether the equation $R_{ik} - \frac{1}{2} g_{ik} R = T_{ik}$ has any reality left within it in the face of quanta. I vigorously doubt it. In contrast, the left-hand side of the equation surely contains a deeper truth. If the equation $R_{ik} = 0$ really determines the behavior of the singularities, then a law describing this behavior would be justified far more deeply than the aforementioned equation, which is not unified and only phenome-nologically justified" (Doc. 348). It is exactly this project that Einstein and Grom-mer believed they had made significant advances on only a few months later: de-riving the equations of motion from the vacuum Einstein equations, without any appeal to an energy-momentum tensor.[11]

Despite the relevance of this result for general relativity proper, Einstein's cor-respondence teaches us why something he had deemed a pseudosolution to the pro-cess of giving an account of matter (considering particles as singularities in the field) was a worthwhile approach to describing the motion of matter in general rel-ativity. He expected that the pseudosolution would point the way to a proper solu-tion: a comprehensive account of (quantum) matter within the realm of classical field theory. He emphasized that hope in the final sentence of the paper, as he had done earlier when writing to Besso and to Ehrenfest (Doc. 450) five days after pre-senting it to the Prussian Academy on 6 January 1927.

Shortly thereafter, Herglotz expressed enthusiasm for Einstein and Grommer's result (Doc. 468). He handed the proofs of the paper to Weyl, who commented in detail, and was less enthusiastic, for he "did not see what in it goes beyond my own development" of this topic (Doc. 473). In his reply, he did not address Weyl's crit-icism that Einstein had failed to acknowledge, or significantly improve upon, Weyl's earlier work on the problem of motion in general relativity (Doc. 514).[12] Instead, he focused on his own motivation for taking up the problem of motion in the first place: the question of whether the "field equations as such are to be con-sidered as falsified because of the facts of quanta—or not." On this occasion, Weyl also took the opportunity of returning to an earlier disagreement, namely, Ein-stein's "measuring-rod objection" to Weyl's unified field theory of 1918 (*Einstein 1918g* [Vol. 7, Doc. 8]). He observed that the new quantum mechanics justified his introduction of the scale factor in that earlier theory (*Weyl 1918*) when reinterpreted as a phase factor, by making imaginary the exponent quantity that depends on the electromag-netic potential. Thus, the scale invariance of Weyl's original theory (from which the term "gauge invariance" is derived) was converted from a state-ment about scale (connected to measuring rods) to one about phase (connected to Schrödinger's wave function).[13] As Weyl observed to Einstein, the theory thus has less to do with field unification than it does with field quantization. It may be argued that this repurposing of the mathematics from Weyl's 1918 theory consti-tutes the inception of modern gauge theory.

In the previous volume we found Einstein already firmly convinced that a satis-factory theory of geomagnetism would have to be connected with a fundamental relation between gravitation and electromagnetism, as his repeated complaints for failing to formulate such a theory testify (see his letters to Auguste Piccard, Ehrenfest, and Kaluza, Vol. 14, Docs. 379, 384, 444). An episode in the present volume allows a glimpse of his effort to forge such a connection.

It was at the meeting of the GDNÄ in Düsseldorf, 19–25 September 1926, that the young physicist Teodor Schlomka approached Einstein about a plan to test whether the geomagnetic field is due to the rotation of the Earth by measuring the second derivatives of the geomagnetic potential, and Einstein talked about his idea of bulk matter behaving like an electrically charged mass having a "ghost charge" and producing a magnetic field when in rotation (see Doc. 442, and Introduction to Vol. 14, sec. V).[14]

Schlomka subsequently initiated a correspondence with Einstein on his planned experiment. He offered to verify Einstein's idea in an airplane and to find out whether and how the direction of the observed geomagnetic field in motion with respect to the Earth deviates from the direction found on the ground. He also enlisted various theories of Ottaviano Mossotti, Friedrich Zöllner, Lorentz, Arthur Schuster, and William Swann intended to explain the Earth's magnetic field by claiming a tiny difference between the absolute charges of the electron and the proton, just sufficient to get a surplus attraction and electric charge, which might account for gravitation and geomagnetism, respectively. Einstein, too, believed that rotation of bulk matter alone could not be the source of the geomagnetic field, since otherwise translational motion should also produce a similar effect, which, however, had not been observed by ships crossing the Atlantic from Europe to America (Doc. 475). He thought that Maxwell's vacuum equations ought to be modified in order to obtain a satisfactory explanation. Schlomka attributed the lack of a translational effect to the ships' low speed during these observations (Abs. 755) and maintained that his measurements in flight had a fair chance of success.

In their further correspondence, and despite Einstein's skepticism, Schlomka developed a detailed program of observations, and discussed the expected results calculated for various conditions (Docs. 483 and 486). He also examined earlier similar experiments by Michael Faraday, Piotr Lebedev, and Harold Wilson (Abs. 781). He attributed the negative results of their attempts to their being carried out at ground level rather than in a moving flight system. Schlomka performed his experiments and reported on them at the end of March 1927 (Abs. 809). He also met Einstein in his home on 1 April, and proceeded to test Einstein's idea in July. But Einstein found the results unsatisfactory.[15]

### III.  Geometrization and Kaluza-Klein Theory

We learn from the correspondence with Rainich that Einstein would consider a unification of gravity and electromagnetism satisfactory only if it could account for the quantum properties of matter. Likewise, we learn from Einstein's correspondence with Hans Reichenbach (Docs. 224, 229, 234, 239, 244, 249) that another aim often associated with Einstein's project of finding a unified field theory—namely, that of "geometrization"—was in fact not one of his desiderata.

This particular part of Reichenbach's correspondence with Einstein was triggered by Einstein's latest attempt at a unified field theory (*Einstein 1925t* [Doc. 17]). In March 1926, Reichenbach (see Illustration 27) wrote that all the recent attempts at a unified field theory felt somewhat "artificial" to him (Doc. 224); possibly to his surprise, Einstein swiftly agreed (Doc. 230). Emboldened, Reichenbach sent a seven-page manuscript (enclosed with Doc. 235) in which he argued that the geometric interpretation of electricity in previously suggested unified field theories was only a "visualization" of the physics, and not itself something physical. He aimed to show this explicitly by providing a reformulation of the coupled Einstein and Maxwell equations, and especially of the Lorentz force law, in which both the electromagnetic and the gravitational field appear equally "geometrical" by being absorbed into a generalized affine connection whose geodesics are to be traced by charged particles.

Einstein's comments on the manuscript engaged with the details, rather than with the message. He had found more than one "fly in the soup." His most important objection was that one affine connection allows only for one type of particle, that is, one ratio of electric charge to mass, to move on its geodesics (Doc. 239). Reichenbach humbly replied that Einstein had misunderstood him as trying to set up a new unified field theory, whereas in fact he had only intended to give a representation of existing physics, one that showed that geometricity (in the sense developed by Reichenbach) is a matter of mathematical representation rather than of physical content (Doc. 244). Einstein could identify with this latter message. He wrote that it "is wrong to think that 'geometrization' is something essential. It is only a kind of crutch for discovering numerical laws. Whether one links 'geometrical' intuitions with a theory is an inessential private matter" (Doc. 249). Thus, Einstein and Reichenbach agreed to agree. But their reasons for the rejection of geometrization as a signal aim of general relativity and unified field theories were rather different.[16]

One of Reichenbach's main observations was that while the metric tensor in general relativity has rods and clocks as its "physical indicators," there is no such indicator for the separately defined affine connection in the attempts at a unified field theory by Weyl, Eddington, and Einstein. Reichenbach now aimed to make charged particles the physical indicator of his generalized connection and to have them move on the geodesics of that connection. Given that Einstein had pointed out that this would only work for one specific charge (such as that of the electron in particular), it is remarkable that he did not compare Reichenbach's approach to Kaluza's attempt at a unified field theory, which allows for *all* charged particles to move on the geodesics of a five-dimensional connection.

Indeed, it seems it was Einstein who had introduced this idea and had suggested to Kaluza on 28 April 1919 that he incorporate it into his original paper (Vol. 9, Doc. 30). But Einstein ended up not communicating the paper to the Prussian Academy for publication, as he had originally offered to Kaluza to do only a week earlier (Vol. 9, Doc. 26). The reason was that Einstein worried about the status of the cylinder condition in the theory and connected to that, the nature of the fifth dimension as compared to the other four spacetime dimensions. Kaluza's original cylinder condition stated that no physical quantity could depend on the fifth coordinate, that is, that all derivatives with respect to the fifth coordinate vanish. Even after having reconsidered that it might have been a mistake not to communicate the paper and having offered again to send it to the Prussian Academy two years later (Einstein to Kaluza, 14 September 1921 [Vol. 12, Doc. 270]), Einstein ended up criticizing the cylinder condition in the first paper he cowrote with Jakob Grommer (*Einstein and Grommer 1923a, 1923b* [Vol. 13, Doc. 12]). This was also the first paper in which Einstein demanded that any satisfactory unified field theory needed to allow for electron solutions. He argued that Kaluza's theory was not fit to accomplish this end.

In the present volume, we find Einstein revisiting Kaluza's theory, especially the interpretation of the cylinder condition, and thus also the status of the fifth dimension. He might have been prompted to do this by Ehrenfest and Lorentz, who alerted him to new work by Oskar Klein, who had likewise—and independently of Kaluza—attempted to use five-dimensional pseudo-Riemannian geometry to produce a unified theory of gravity and electromagnetism (Doc. 302 and Abs. 506). But Klein was even more ambitious than Kaluza in that he attempted not only to unify gravity and electromagnetism, but also incorporate Schrödinger's newly found wave function within this framework.

Both Ehrenfest and Lorentz urged Einstein to come to Leyden and join their meetings with Klein. Einstein wrote that he had to finish some things before vaca-tioning with his sons (Doc. 319), but asked to see Klein's paper two and a half months after he had been first informed of it (Doc. 356). Most likely at Ehrenfest's suggestion, Klein wrote to Einstein directly in late August 1926 (Doc. 363). He sent Einstein not only the manuscript of his paper (*Klein, O. 1926*), but also proposed how he wanted to develop the theory further. In particular, he explained his idea of assuming a periodicity of the fifth coordinate, averaging over it, and reconceptualizing Schrödinger's wave function as a component of the five-dimensional metric by effectively dropping the sharpened cylinder condition and using the same version of the condition that Kaluza had used seven years earlier.

A few months later, Einstein wrote two papers on Kaluza's theory, *Einstein 1927i* (Doc. 459) and *Einstein 1927j* (Doc. 480). In the first paper, Einstein focuses on his old concern regarding the meaning and consequences of the cylinder condition. First, he gives a coordinate-independent formulation of both the original cylinder condition (used by Kaluza) and of the sharpened cylinder condition (initially used by Klein). The latter is both more, and less, restrictive than Kaluza's condition: it is less restrictive in that it only demands that the derivatives of the metric with respect to the fifth coordinate, rather than of all physical quantities, vanish; but it is also more restrictive in that it demands that the component $g_{55}$ be a constant, rather than a variable, as permitted by Kaluza. Second, Einstein identifies the invariants of the reduced group of five-dimensional coordinate transformations resulting from the sharpened cylinder condition, namely, a four-dimensional symmetric tensor and a four-dimensional antisymmetric tensor, both of second rank. And third, he argues that thereby "Kaluza's idea offers a deeper understanding of the fact that besides the symmetric metric tensor ($g_{mn}$) only the antisymmetric tensor ($\phi_{mn}$) of the electromagnetic field (which is derivable from a potential) plays a role" (Doc. 459).

In the second paper, Einstein examines the relation between general relativity and Kaluza-Klein theory. He investigates how the five-dimensional metric should be projected into four dimensions so as to recover the original Einstein-Maxwell equations exactly, and notes that Kaluza had only managed to derive them approximately. He concludes by stating that in order to do this "in the usual form, one must presume the 0-direction to be *spacelike*" (Doc. 480). Both Kaluza and Klein had assumed the fifth dimension to be spacelike; Einstein believed he had found a consistency argument in its favor. It is noteworthy that Einstein did not comment at all on the links that Klein had tried to forge between five-dimensional general relativity and Schrödinger's wave mechanics.

The middle of the second paper consists of a thorough examination of how five-dimensional geodesics are related to four-dimensional ones. This had been investigated in even more detail by Vladimir Fock (see *Fock 1926*), whose student, Heinrich Mandel, was working with Einstein in Berlin at the time. In an addendum to the two papers, Einstein wrote that Mandel had informed him that everything he had done in the paper was already contained in *Klein, O. 1926*, and that *Fock 1926* should also be consulted. In the draft of the addendum, much longer than its published version, Einstein acknowledged Klein, Fock, and also Mandel, whom he credits with having thought of the five-dimensional approach independently of Kaluza. It is unclear why Einstein eventually deleted these assertions of co-priority, or the reference to Mandel. Whatever the case may be, even before the publication of his two papers on Kaluza-Klein theory, Einstein, in response to a request, recommended Kaluza as successor to Gerhard Kowalewski, Professor of Mathematics at Dresden Technical University (see Abs. 621 and Doc. 408). He also highly praised and recommended Kaluza in a letter to Karl Herzfeld shortly before presenting his own first paper on Kaluza's theory to the Prussian Academy (Doc. 447).

## IV.   Family Life

Einstein's family-related correspondence during the period covered by this volume is substantial, amounting to a fifth of all letters. At the forefront of his concerns at this time were his sons, Hans Albert and Eduard, albeit for different reasons (see Illustration 12).

In the previous volume, disagreement over a distribution of funds from the investments of the Nobel prize money had led to Einstein's temporary estrangement from Hans Albert in early summer 1923.[17] That rift proved to be merely a prelude to a more severe discord that began in the fall of 1925, after Hans Albert expressed his intention to enter upon a permanent relationship with Frieda Knecht, a former Zurich neighbor (see Illustration 15). Einstein believed that his twenty-one-year-old son suffered from strong inhibitions toward women. He intended to instruct him "inconspicuously" (Docs. 7 and 45) and hoped that Hans Albert would thereby abandon plans for a marriage that would be a "pity for the good breed!," as he wrote to his first wife, Mileva Marić (Doc. 63). His son had spoken "with great enthusiasm about marriage and argued against the importance of making sure of good breeding" (Doc. 79). Einstein hoped that the "rather dangerous" situation would eventually resolve itself through patience (Doc. 89).

But by October 1925, Einstein and Marić decided that action ought to be taken. Einstein pinned his hopes on "a good-looking woman in her 40s," whom Hans Albert had met, who might distract the boy from his infatuation, and recommended that Mileva send him to Berlin for a year of study during which he might be "cured" (Doc. 88). But only a day later he retracted this suggestion after having discovered that the woman in question was a friend of Mileva and merely had "a human interest" in their son.

It was both Knecht's pedigree and her age that worried Einstein, with consequences he thought "too terrible" to contemplate in light of "unfavorable hereditary factors" present, especially on the mother's side (Docs. 95 and 104). The age difference was to him "even worse" than in the case of his own first marriage (Doc. 105).[18] For Einstein, "the drama" of his marriage at a young age was now being repeated, as in the biblical saying to which he alluded: "visiting the iniquity of the fathers upon the children" (Doc. 89). The "iniquity" in question was presumably Einstein's conflict with his own parents over their fierce opposition to his plans to marry Mileva[19] and the alleged genetic "inferiority" that Mileva, and to a lesser extent he himself, had brought into their union.[20]

Einstein even asked friends to intervene (Docs. 89 and 99). Hermann Anschütz-Kaempfe, who had intermittently employed Hans Albert at his factory in Kiel, played a sizable role in these endeavors: after meeting with Knecht, Anschütz-Kaempfe concluded that she was "a psychopath who has pathologically exaggerated egocentricity," "decidedly degenerative characteristics, dwarfish stature & nascent development of a goiter," and "the skull formation [is] also pathological" (Doc. 110). He counseled removing Hans Albert, whom he deemed socially isolated and financially hampered, from Frieda's influence (Doc. 111). While conceding that his son had suffered a lot because of family circumstances, Einstein believed that it was primarily his son's personality and external projection of his woes that had brought about this situation (Doc. 105). He also blamed Mileva: had she sent their son to study in Munich instead of enabling him to remain in Zurich, the boy would have gained a better insight into women and human relationships (Doc. 195).

Einstein's worries gradually escalated. He came to believe that it would be "a crime" for Hans Albert to have children with Knecht, and that everything should be done to avert "a catastrophe" (Docs. 135 and 185). He now took the radical step of completely breaking off ties with his son, telling him not to write or visit until he had "resolved this conflict" (Doc. 286). Although Mileva and Anschütz-Kaempfe both thought it unwise to make Hans Albert choose between his love interest and contact with his father, Einstein did not relent (Docs. 191, 195, 263).

He also set his hopes on a senior psychiatrist in Zurich (Doc. 211) and asked his friend Heinrich Zangger to intervene. He believed Knecht's mother to be in an insane asylum and was "horrified by the thought of offspring" (Doc. 243). Zangger discovered that the mother was suffering from an autoimmune disease and possibly from depression as well (Doc. 312). By May 1926, Einstein feared that Hans Albert had "fallen into a kind of servility to the girl" and entertained slim hopes that "the disaster" could be staved off (Doc. 286). By June 1926, the increasingly stressful situation manifested itself in physical symptoms, and Einstein informed Mileva that he needed to convalesce in the mountains (Doc. 309).

But in July, following intense discussions with Hans Albert while on a visit to Zurich, Einstein wrote that the liaison was "a real 'relationship'," thereby implying that it was intimate in nature, and expressed his approval. A deterioration in Mileva and Hans Albert's rapport reminded Einstein again of his own past (Docs. 325 and 328), and by the fall he was again pessimistic and asked Eugen Bleuler, the prominent Zurich psychiatrist, to intervene (Doc. 366). Bleuler concluded that there was "a presence of parallel genetic predispositions in the two families." He advised against the marriage (Doc. 382) but failed to dissuade Hans Albert. Einstein concluded that "nothing more can be done" (Doc. 397).

In January 1927, Hans Albert, who had recently completed his engineering studies at the ETH, wished to visit Einstein in Berlin and seemed to be open to accepting his father's assistance in finding employment (Doc. 449). Eventually, he obtained an appointment at a steel fabrication company in Dortmund without his father's help (Doc. 466). A visit with Einstein prior to taking up this new position (see Illustration 14) led to an amelioration of their discord (Doc. 467). Einstein softened his stance and declared that if Hans Albert resolved not to have any children with Frieda, he would resign himself to their marriage, and yet he vehemently objected to Hans Albert's imminent plan to move in with Knecht in Dortmund (Docs. 469 and 474).

The correspondence with Hans Albert in this volume ends on a sour note: Einstein demanded that his son never bring Frieda to Berlin, because he "could not bear it," and issued a dire prediction that the day would come when Hans Albert would want to separate from her: "It all results from the fact that *she* was the first to grab hold of you and you now view her as the embodiment of all femininity. We all know how quixotic people succumb to fate" (Doc. 484). Intriguingly, there is no mention in the correspondence of the actual marriage ceremony, which took place in Zurich on 17 May 1927.[21]

At the beginning of this volume, Einstein's younger son Eduard was almost fifteen. Over the next two years, his intellectual development, as reflected in the correspondence with his father, is truly remarkable (see Illustration 13). Their exchanges reveal Eduard's increasingly probing mind and, at times, agonized self-analysis, and Einstein's obvious delight at his son's intellectual growth and deep concern for the boy's emotional resilience.[22]

After Eduard's visit in Berlin in July 1925, Einstein expressed great pleasure

(Doc. 25). And while he thought that Eduard might be intellectually more gifted than his brother, he also characterized him as being egotistic, too ambitious, and lacking emotional balance and close contact with others. He found Eduard to exhibit feelings of isolation, anxiety, and other inhibitions, and believed him to take "a lot" after Einstein himself. He envisaged that the boy's life would not be easy (Doc. 45). Einstein derived much pleasure from Eduard's many poems (Doc. 78), but was concerned with the boy's "delicate nervous system." He urged Mileva to make sure that Eduard not become too lonely so as not to follow his brother's fate (Doc. 185). But when Eduard sent him an exacting self-assessment in April 1926, Einstein now "felt like a hen that has hatched a duck egg." While he really enjoyed Eduard's sincere and comical reflections, he warned him of the pitfalls of taking oneself too seriously (Doc. 257). He deemed it important to spend as much time as possible with Eduard during this period of "stormy development" (Doc. 309).

During an intense exchange of letters in the fall of 1926, Einstein confided that Eduard's letters reminded him of his own adolescence, and recalled having similarly alternated between despondency and self-confidence. He counseled that youthful "heroism" needs to be ameliorated by humor and by meshing "into the social engine." He tried to allay Eduard's pessimism and nihilism, rooted in fear of worthlessness, and assured his son that he brought him "great joy" since he did not "go through life apathetically but rather as a seeing and thinking being" (Doc. 415). He was "joyous like a child with the bottle" when a letter arrived because he saw Eduard "agonize about the principal things in life" (Doc. 434).

Einstein shared Mileva's concern that, because of Eduard's success as a writer, it was "dangerous for him if one courts him too much." In Einstein's mind, it would be "ruinous for him if his ambition is stirred up," since Eduard could lose "the contemplativeness without which deeper development is impossible" and might become embittered later in life. The boy should therefore be strongly encouraged to pursue a "normal career that will give him a certain security of social status which will ensure his internal equilibrium." Creative literary work as a primary occupation was for Einstein "an absurdity" (Doc. 488).

Eduard's self-perceptions also underwent many changes, as he himself recognized. In February 1926, he thought his predispositions guided him toward intellectual rather than emotional art, and not only in music (Doc. 190). Two months later, he delivered a quite critical, and quite exquisite, self-portrait: he was "generally fickle and erratic in character," and egoistic. He was both "tremendously lazy," but also disapproving of his laziness. In his mind, he was made up mainly of such "dual personalities" (Doc. 241). Half a year later he felt indifference to matters that a few years earlier were his "supreme goals." Yet he acknowledged that no new aspirations had taken their place.

Eduard's comments on his father reveal how very difficult it was for him to be the son of a celebrity. In February 1926, he expressed admiration for Einstein's ability to "always have a suitable maxim at hand" (Doc. 190). He was unsure of the sincerity of Einstein's praise for his poems (Doc. 241), and reached the conclusion that he had to "be careful" vis-à-vis Einstein, as his father was "superior" to his "primitive thoughts." He confessed that at times it was pleasant to have such an "illustrious father," but at other times "it is rather uncomfortable... One feels so insignificant" (Doc. 274). Half a year later, in reaction to Einstein's attempts to normalize and contextualize his son's agonized introspection, Eduard concluded that it was "completely hopeless to argue" with his father, as Einstein himself had already thought about these matters and was "at least 30 years ahead of him in every respect." He also decided that it was "very, very dangerous" to present Einstein with observations that were not completely "nailed down." Ironically, both father and son feared each other's criticism (Docs. 195 and 433).

Eduard's letters gradually became more intellectual in both tone and content. He discussed simplicity and feeling in art, and defended Schopenhauer, the adolescent Einstein's erstwhile hero, against his father's criticism (Doc. 190). Between May and December 1926, father and son engaged in their most heated discussions to date. Eduard began to express beliefs that may well have been intended to provoke. In his opinion, human achievements, especially those of individuals, were "completely insignificant and indifferent," and the importance of the mind was "overrated" (Doc. 274). He pursued this line of thought by arguing that there was "a desperately small difference between a genius and an idiot." Works of art were lacking "any intrinsic value" and, indeed, science was "totally useless" (Doc. 414). Einstein strongly disagreed "about the worthlessness of intellectual production." It was impossible not to acknowledge the highest stage of consciousness as "the most supreme ideal," he wrote. He believed that Eduard was advocating eudaimonism, which Einstein described as "a dreary swine-herd ideal." In contrast, "cognition in the artistic and scientific sense [was] the best thing we have." He suggested a palliative to what he perceived as Eduard's nihilism: the boy should "become a small cog in the great machinery of life: If one hears the angels singing a few times in one's lifetime, then one can give something to the world and one is a particularly happy and blessed person" (Doc. 415). But Eduard did not consider eudaimonism to be such a dismal ideal. To him, science was "harmful" because of its overemphasis on cognitive activity. Those who follow mostly intellectual pursuits "sire sickly, nervous, and sometimes completely moronic children," Eduard wrote in typically self-denigrating fashion, citing himself as a prime example (Doc. 433).

By mid-1925, Einstein's relationship with Mileva Marić had begun to improve.

Einstein was looking forward to lodging in her newly purchased house when he next visited Zurich,[23] and also proposed that Mileva and Eduard visit him in Berlin (Doc. 7). But by the fall of 1925, when Mileva complained about his past behavior toward her and informed him that she intended to write her memoirs for publication, Einstein reacted with vehement opposition and ridicule: "Does it not enter your mind at all that no one would care less about such scribblings if the man that they were about had not, coincidentally, accomplished something special? If someone is a nobody, there is nothing to object to, but one should be truly modest and keep one's trap shut. This is my advice to you" (Doc. 95). Mileva replied without further acrimony, and Einstein's ire was allayed (Doc. 99). In anticipation of his second stay at her house, he proposed that they vacation together with Eduard in the Swiss Alps. He was indifferent to public opinion of such an arrangement "due to complete desensitization" (Doc. 309). Their united opposition to Hans Albert's plans to marry Frieda Knecht had evidently fostered greater amiability. Indeed, by early 1927, he was reassuring his former wife that she would gradually realize that "there is hardly a more pleasant divorced man than I" (Doc. 448). He was also pleased that Mileva and their sons were "no longer so hostile" toward his second wife Elsa Einstein (Doc. 485).We do not know whether Elsa was comfortable with his lodging arrangements in Zurich in the summers of 1925 and 1926, but Einstein seemed to be rather oblivious of its effect on her. In a postcard to Elsa from Zurich he informed her that he was "sitting with the boys and your predecessor" (Doc. 325).

During the period covered by the previous volume, Einstein's marriage to Elsa had been sorely tested by his liaison with his secretary Betty Neumann, the first documented extramarital affair of his second marriage.[24] After his return from South America, and possibly in response to pressure from Elsa, Einstein hired a male secretary, Siegfried Jacoby.[25] A few months later, while lodging with friends in Düsseldorf, Elsa apparently reacted with jealousy to the way Einstein described his hostess. Einstein replied insensitively: "What funny business are you writing there about Mrs. Lebach? You think that I would be capable of being disloyal in such a way to a splendid man whose hospitality I was enjoying and with whom I was socializing as a friend?" (Doc. 376).

The year 1926 was very difficult for Elsa, who lost both parents within six months (see Illustration 10). In August, Einstein decided to return home a few days early to provide "wife no 2" with moral support (Doc. 348). In a rare unguarded passage in which he disclosed his perception of Elsa to Hans Albert, he admitted that "although she can sometimes get on one's nerves and is no great intellect, she excels in kindheartedness" (Doc. 474). Einstein's letters reveal other questionable views on women. He described the physics student Esther Polianowsky-Salaman, for whom he had written a recommendation, as "a hussy, i.e., [someone who] replaces strength with cunning, relies on her attractive appearance" (Doc. 100). And while he praised Tatiana Ehrenfest's extraordinary intellect, he then added: "If she were a man, something significant would come of it. However, I believe she will not muster the energy for that" (Doc. 376). He also made critical remarks on American women, similar to those that had caused him trouble in 1921.[26] In a statement defending Charlie Chaplin's right to privacy in his divorce case, Einstein commented that "in Europe the petticoat rule is not so strong" (Doc. 481). And in a rather callous letter to Hedwig Born, he belittled her, and other women's, lack of creativity (Doc. 444).

## V.   A Bet on Relativity: Miller's Ether Drift Experiments

Einstein first heard of Dayton C. Miller's experiments on ether drift while visiting Princeton University in May 1921 and shortly thereafter met him at the Case School of Applied Science in Cleveland, where Miller was a professor (see Illustration 30).[27] It was at Case that Albert Michelson and Edward Morley had earlier conducted the famous experiments on ether drift. Beginning in 1900, Miller and Morley had published papers confirming the null result of the Michelson-Morley experiment (*Morley and Miller 1905*). Like many physicists at the turn of the century, Michelson, Miller, and Morley were firm believers in the ether theory of light. But unlike others, they retained a preference for the ether theory for many years afterward. Miller certainly continued to direct his experimental program within this framework, and had become convinced that the null result of the ether drift observations was due to ether drag and the "heavy stone walls of the building within which the apparatus was mounted." He accordingly had set up the apparatus "on high ground near Cleveland, covered in such a manner that there is nothing but glass in the direction of the expected drift" (*Morley and Miller 1907*).

Einstein could not examine the apparatus in 1921, since by then it had been moved to Mount Wilson in California, at the invitation of the observatory's director, George Ellery Hale (*Swenson 1972*, p. 192). Nevertheless, Einstein believed that Miller's subsequent results were due to a failure to control adequately the temperature in the vicinity of the instruments. Miller's desire that his apparatus should be, as far as possible, open to the ether wind, tended to render it also unusually open to the elements, especially sunlight.[28]

During their encounter in 1921, Einstein and Miller had discussed a feature of the ether drag theory that would play a central role in the 1925 debate between their respective supporters that is documented in the present volume.[29] If the solid mat-ter of the Earth completely drags the ether along with it, so that no ether drift could be detected at the surface of the Earth, still it must be true that in space far from the Earth the ether must be unaffected by the Earth's motion. Accordingly, it was natural to suppose that a gradient must exist, so that, at sufficiently high altitudes above the Earth, some ether drift ought to be measurable. At the time, Miller had already begun a new series of experiments in a lightly constructed building at Mount Wilson, using substantially the same apparatus. It was there that he claimed to have detected a positive ether drift, in contrast to the earlier null result in the Cleveland basement. He always claimed, however, that there had been a similar, but smaller, positive effect in the slightly elevated location in Cleveland.

The results of Miller's 1921 experiments were published the next year (*Miller 1922*) and reached Einstein through a letter from Max Born of 6 June 1922 [Vol. 13, Doc. 320]), who commented: "The Michelson experiment belongs to things that are 'practically' a priori; I believe not a single word of the rumor." Einstein took a similar view. Over the next three years Miller published only brief reports. By 1925, finally convinced that he was consistently seeing an ether drift of some kind on Mount Wilson, he published at greater length and in multiple places.

It was the editor of the journal *Science,* Edwin E. Slosson, who sent Einstein the proofs of two papers by Miller in June 1925 (*Miller 1925a* and *1925b*), asking for Einstein's opinion (Doc. 12). Einstein replied cautiously: experiment being the supreme judge, he was awaiting more complete details (Doc. 13). The same reserve can be seen in a letter to Robert A. Millikan: "Miller's experiments rest on sources of error. Otherwise the entire theory of relativity collapses like a house of cards" (Doc. 20).

Most physicists expressed serious doubts, first among them Millikan and his staff at Caltech, who were in a position to see the apparatus on Mount Wilson. Already in July 1925 they launched a "counterattack," both on the heights of Mount Wilson and on the lower level of Caltech's laboratories, because, as Paul Epstein wrote to Einstein, "our circle accepts Miller's somewhat daring statements with great reserve; we hope we will be in the position to check his measurements by other observers with other kinds of instrument" (Doc. 31). Upon Einstein's inquiry as to whether Millikan and Epstein were intending to "gather new evidence on the problem of 'ether drift'" (Doc. 58), Epstein disclosed that they intended to repeat Michelson's classical experiment with a modified apparatus and that the idea belonged to the young student Roy Kennedy (Doc. 72). A year later, Epstein reported that the "repetition of the Michelson experiment by Kennedy had led to a completely negative result, even though the sensitivity of the apparatus was four to five times higher than that of Miller's" (Doc. 372 and Abs. 662). Epstein's impression had grown more and more certain that "the whole story is a humbug," and that Miller did not understand his results and had drawn conclusions not supported by his own data. In addition, Miller apparently changed his interpretation fortnightly (Doc. 72).

Born was also an eyewitness to Miller's apparatus, and, as Hedwig Born wrote to Einstein, was "horrified by the mess of the experimental arrangement" (Abs. 408). Michelson, too, was at Mount Wilson and Caltech at the time, and, as Epstein put it, was "very reserved, of course. The only thing I have got from him is that in his opinion Miller does not vary sufficiently the circumstances of the experiment" (Doc. 72).

Several of Einstein's correspondents commented on the implications of Miller's results for astronomy. Most prominent was the Eddington's statement that the experience of astronomers with stellar aberration invalidated Miller's claims (*Eddington 1925c*). Stellar aberration had played a key role in Einstein's arguments in favor of the theory of relativity, since it shows a definite effect of the motion of the Earth relative to the starlight's source on the direction of the propagation of light. Miller claimed that he could not confirm Michelson and Morley's experiment at the high altitude of Mount Wilson. If this experimental result depended upon altitude, Eddington demanded to know why observatories on mountaintops did not report any difference in stellar aberration from that observed at sea level.

Miller's preferred explanation for his results was some form of ether drag theory, as argued by Ludwik Silberstein (*Silberstein 1925c*). A clear implication of this interpretation would be that his observations on Mount Wilson were successfully measuring the solar system's motion through the luminiferous ether. Einstein himself regarded this as an interesting aspect of the experiment, commenting to Lorentz that it looked as if Miller's data showed a constant direction against the fixed stars, though he doubted that this could be due to inertial motion. Nevertheless, "if this is confirmed, then something fundamental lies behind it. Planck and Laue view it very skeptically" (Doc. 310).

Emil Cohn criticized Miller for his refusal to state what motion of the solar system his results were actually consistent with (Doc. 64). Several of Einstein's correspondents were suspicious of Miller's failure to make a bold statement as to the velocity and direction of the Earth through space, including not only the Earth's rotational and orbital motion within the solar system, but also the solar system's motion through interstellar space. Miller merely stated in his 1925 papers that he was working on calculating such a quantity. Some scientists, such as Cohn, evidently suspected Miller of being unwilling to let his own theoretical model undergo a possibly falsifying test.

As it happened, astronomers' understanding of the motion of the solar system was undergoing a major transformation at this time. Previously it was believed that the solar system was moving within the system of nearby stars at a velocity of 20 km/s in the direction of the constellation Hercules, toward a point known as the solar apex. This was discovered in the eighteenth century by William Herschel, based upon small changes in the position of nearby stars, relative to earlier observations by the first Astronomer Royal, John Flamsteed (*Herschel 1783*).[30]

The motion toward Hercules and Miller's claim that even the walls of a building might be sufficient to drag ether along with the apparatus undoubtedly inspired Auguste Piccard's plans to, as Einstein put it, "Miller-in-a-balloon" (Doc. 85). Piccard hoped to time the season of his measurements, and conduct them at night, in order that the three known components of the Earth's motion (rotational, orbital, and toward the solar apex) would all be roughly parallel, thus giving the largest possible motion with respect to the putative ether (Doc. 74). He detected no such motion, in agreement with Einstein's relativity postulate. As Miller tried to measure the absolute velocity of the solar system, astronomy was undergoing a revolution that revised the estimate of this velocity upward by an order of magnitude.

Coincidentally, a major contribution to this revolution was presented by Gustaf Strömberg, an astronomer at the Mount Wilson Observatory, at the same session of the National Academy of Sciences in 1925 at which Miller announced his results. Josef Weber, who had shown that Miller's results exhibited a minimum ether drift when they should have shown a maximum—when the Earth's motion was oriented roughly toward the solar apex (Abs. 188 and *Weber 1926b*)—informed Einstein of Strömberg's results, which showed that the solar system moves at a velocity of some 300 km/s with respect to certain old stars (now called halo stars), globular clusters, and other galaxies (*Strömberg 1925*). The best interpretation was, and is, that the solar system orbits the center of the galaxy at this velocity and that the earlier observed velocity of 20 km/s is merely the velocity with respect to nearby stars that share the solar system's orbital motion about the galaxy's center. This lower velocity was still expected by experimenters like Piccard (Doc. 74) who were not, like Miller, in the fortunate position of conducting their experiments on the grounds of the world's leading observatory. Strömberg's results, together with work by the Dutch astronomer Jan Oort and the Swedish astronomer Bertil Lindblad, established our current picture of the structure and dynamics of the galaxy. Since Strömberg was at Mount Wilson, he was eventually successful in pinning Miller down to a specific claim of the solar system's motion through space: 200 km/s in a direction about 23 degrees away from the solar motion in the galaxy (*Strömberg 1926*). This was close enough to Strömberg's conclusions to put new heat into the reception of Miller's work.

Miller's claim that his results agreed with the exciting discovery of the Earth's motion about the Galactic center caught the imagination of many senior physicists, as suggested by correspondence between Wilhelm Wien and Schrödinger. Their lengthy discussion on the topic began in mid-September 1925, when Wien asked Schrödinger "what do you think about the positive result of the Michelson experiment?... The observations on the shift of the interference patterns reproduce the motion of the solar system in the universe, in agreement with the latest observations of Strömberg!! It is the most astounding result in physics." Their exchanges were animated by their efforts to organize German and Swiss experiments designed to replicate Miller's results. One even detects a certain degree of satisfaction in their anticipation of the forthcoming counterrevolution in physics. Wien wrote: "If the observations are substantiated—as can hardly be doubted any more—the theory of relativity, the special as much as the general one, is finished, and we must go back again to our old ideas of 25 years ago." In his reply, Schrödinger commented on "the hardly noticeable 'counter-propaganda' in the Jewish circle of physicists" trying to play down the result, which he decried as unfair to Miller (*Mehra and Rechenberg 1987*, p. 453). Intriguingly, Schrödinger, who played an active role in organizing facilities for the Swiss-based replications, never mentioned the Miller experiments or these replication efforts in his correspondence with Einstein in this volume, which occupies the same few months as his correspondence on the subject with Wien.

It was this triumph of Miller's that set in motion the most important of the replicating experiments. At the end of 1926, Hale informed Einstein that Michelson himself would perform an interferometer experiment atop Mount Wilson (Doc. 425). It would take several years before results were announced, but Michelson firmly ruled out any inertial motion of the Earth, even a tenth as great as demanded by the astronomers, whose conclusions were by then firmly established (*Michelson, Pease, and Pearson 1929*).

A key aspect of Miller's success lay in convincing scientists that his work agreed with Strömberg's startling new discoveries. In reality, his interferometer measured, at best, a velocity an order of magnitude less than what Strömberg had seen, but Miller could reconcile them by his own determination of the Earth's direction of motion in the galaxy (*Miller 1933*), which did not accord with subsequent astronomical findings. Furthermore, as pointed out by André Metz (Doc. 157), Miller's favored ether drag hypothesis was invalidated by Michelson's large-scale version of the Sagnac experiment, which conclusively proved that interferometry could detect the rotational, non-inertial, motion of the Earth (*Michelson and Gale 1925*). Only the theory of relativity could explain the success of the Michelson-Gale experiment combined with the null result of the Michelson-Morley experiment.

One other astronomical result played a role in the controversy, pointed out by Eddington and mentioned by Slosson to Einstein (Doc. 12). At the same remark-able session of the National Academy of Sciences at which Miller and Strömberg presented their results, Hale reported on work by Walter S. Adams, an astronomer at the Mount Wilson Observatory, that demonstrated the existence of a gravitational redshift in the spectrum of the white dwarf star Sirius B (*Adams 1925*). This result was in agreement with earlier calculations by Eddington (*Eddington 1924*). Skepticism about the "third test" of relativity had played, for a time, a role in sustaining the critics of relativity (see, e.g., Vol. 9, Introduction, pp. xxxvii–xl). Eddington, who had also been contacted by Slosson, replied that Miller's attempt to reopen the debate on special relativity came just as the empirical verdict on general relativity was beginning to look quite settled (see Doc. 12).

Einstein suspected that small differences between the arms of the interferometer were to blame for Miller's results: "Temperature differences in the air between the two beams, of the order of magnitude of 1/10 of a degree, would suffice to produce the whole fuss," he wrote to Ehrenfest (Doc. 71). He did not find any reference in Miller's publications as to how this error had been avoided, and suggested as much to both Miller and Piccard (Doc. 219). Miller agreed with the remark that 1/10 of a degree difference would produce the observed displacement of interference fringes, therefore "very elaborate precautions have been taken to eliminate such an effect of temperature" (Doc. 289). However, these precautions had to compete with Miller's desire not to enclose his apparatus in stone walls.

Einstein told Millikan privately that he distrusted Miller's result but had no right to say so in public (Doc. 58). The data reaching Einstein strengthened his opinion, expressed much earlier to Ehrenfest: "Basically I think nothing of Miller's experiment, based on my malicious soul, but I must not say it aloud." He characteristically invoked the Creator, whom he credited "with more elegance and intelligence than that" both to an old friend (Doc. 26) and to Ehrenfest: "the difference between Cleveland and Mount Wilson cannot be so significant, considering the grand scale on which the Old One created the world" (Doc. 49). When he had first learned of Miller's experiments in 1921, he had expressed his distrust in a similar vein: "Subtle is the Lord, but malicious he is not" (see Vol. 12, Introduction, p. liii). Now we see exactly what he had in mind.

Scientists decided upon new observations aimed at replication of results, although Einstein even here had doubts, desperately noting the expense involved: "What can one do now to bring some order to this epidemic? It would be a pity to spend too much money on this shady matter" (Doc. 86). Many of those planning such replication endeavors communicated with Einstein, some seeking help in raising funds. While Kennedy's experiment was certainly influential in turning opinion against Miller, Piccard's balloon-based experiment additionally proved to be of great significance for the history of aeronautics. Although balloons had been used in science previously, and although Piccard himself was an experienced aeronaut, his success in helping rebut Miller encouraged him to combine his physics career with his passion for ballooning (see Illustration 28). Significant funding was needed for manned scientific ballooning, and Piccard saw the Miller affair as a perfect opportunity to interest potential funders, especially with Einstein's "moral support" (Docs. 74 and 87). His brother (and fellow aeronaut) Jean recalled the moment when Piccard resolved to go higher than anyone had gone before: "My twin brother, Auguste, first discussed this proposed flight with me in 1926. He wanted to go to a greater height, not to establish an altitude record, but to determine, if possible, the action of cosmic rays and their quality and intensity at different altitudes." (*Ziegler 1989*, p. 960). Piccard was anxious to contribute to the debate on the origin of cosmic rays, which were widely thought to be of extraterrestrial origin, but which Millikan had argued were terrestrial. Since Millikan's unmanned balloon flights had provided controversial evidence for a noncosmic origin, Piccard argued that a manned balloon flown to high altitudes would facilitate a more reliable experiment. Accordingly, he invented a sealed capsule that enabled him to set an altitude record in 1931 by becoming the first person to ascend to the stratosphere, nearly 16,000 m up. The resulting fame saw him immortalized in the character of Professor Calculus by his fellow Brussels resident Georges Remi, better known by his pen name of Hergé, in the pages of *The Adventures of Tintin*.

Some of those attempting to replicate Miller's results shared his theoretical preference for a result falsifying the theory of relativity. Among them was Rudolf Tomaschek, a student and then assistant of the anti-relativist Phillip Lenard, who hoped for a positive result when, as Einstein put it, he went "to Miller around" on the Zugspitze, in southern Germany (Doc. 85). Tomaschek figured in the discussions between Wien and Schrödinger (*Mehra and Rechenberg 1987*, p 453.). Schrödinger proposed a Michelson-type experiment to be performed on the Jungfraujoch, a saddle ridge between two peaks in the Swiss Alps considerably higher than Mount Wilson, but accessible via railway, and boasted a research station and observatory near the top, which were then under construction. He asked Wien to suggest a good optics specialist to perform the experiment, and Wien nominated Tomaschek, who was experienced in interferometry and had already performed experiments on the very same mountain ridge as part of Lenard's research program. According to Schrödinger, the Swiss were skeptical of Tomaschek's impartiality, given his connection with Lenard, and he was rejected, much to Wien's disappointment. Ironically there is evidence that neither Lenard nor Tomaschek placed much faith in the reliability of Miller's work. All of Tomaschek's experiments indeed contradicted Miller's findings (Doc. 260).

Nevertheless, the Swiss, led by Edgar Meyer, selected Georg Joos for the Miller experiment on the Jungfraujoch. Joos diplomatically enlisted Tomaschek as a partner, proposing that he would do the interferometry and Tomaschek would perform a Trouton-Noble experiment in the new research station (Tomaschek had already performed just such an experiment on the mountain, inside a wooden construction perched high up under a cliff wall). Joos enlisted the Zeiss company of Jena for his optics. He did not believe in Miller's results and doubted the need for replication but, like Piccard, saw the opportunity to acquire funding and new equipment, given the publicity surrounding the Miller affair (Doc. 280). Though Tomaschek did continue his work on the Jungfraujoch, Joos never did bring his instrument up the mountain and instead performed the experiment in Jena. As he put it, echoing Einstein's concern at throwing money away uselessly, "one may rightly ask whether… in view of the financial calamity of German science the expenses for such an expedition could still be justified" (*Mehra and Rechenberg 1987*, p. 458). He pointed out that Miller had eventually rescinded his claim that his effect was in any way connected with the altitude of the Mount Wilson Observatory. Joos eventually placed an upper limit of 1.5 km/s on the ether wind at Jena (*Joos 1930*).

As we have seen, at the outset of the controversy Einstein felt unable to publicly air his skepticism of Miller's findings. Gradually, however, he was encouraged by events to express his doubts more freely. He was even quoted in the press advising the public not to bet on confirmation of Miller's results (Doc. 161). At the end of 1926, and also of the present volume, he took a public stance on the matter. In a short paper in a popular scientific journal, he first summarized the improvements Kennedy and Piccard had made to their instruments compared to Michelson's original one. Even though they could not eliminate completely the disturbing effect of environmental temperature, Kennedy's and Piccard's results disproved Miller's main statement, namely, that there is a drag of the ether by the Earth that changes with altitude. Einstein concluded with a chivalrous funeral oration for the initiator of the debate: "No doubt, it was Prof. Miller's outstanding merit that he initiated a meticulous reexamination of Michelson's important experiment" (Doc. 478).

### VI. German Politics and European Rapprochement

Germany continued to face considerable political turmoil during the period of this volume. Its foreign affairs were dominated, first, by the Locarno Treaties of October 1925 that had been negotiated on Germany's behalf by foreign minister Gustav Stresemann and which guaranteed nonaggression and normalization of relations with Allied Powers in the postwar era.[31] The ongoing negotiations on Germany's entrance into the League of Nations were the second major issue in foreign policy. Following many delays, Germany officially joined the League on 10 September 1926.[32]

Chancellor Hans Luther's cabinet was plunged into a severe crisis days after the Locarno Treaties were signed owing to vehement opposition by German nationalists. The exit of the German National People's Party from the government led to the resignation of Luther's first cabinet; his second cabinet lasted only a few months.[33] Another major issue on the domestic political scene was the struggle over the compensation of former German princes for their confiscated property. After an intensive public campaign, a referendum on the issue failed to obtain the necessary votes in June 1926.[34]

The economic conditions in Germany deteriorated substantially. From mid-1925 onward, a "severe stabilization crisis" set in, caused by the Reichsbank's deflationary policy and a significant decrease in foreign loans, leading to bankruptcies and a steep rise in unemployment. By January 1926, the unemployed numbered more than two million.[35]

Einstein suffered no small amount of soul-searching in deciding what causes to support or endorse. In June 1925, the German pacifist Otto Lehmann-Russbüldt asked Einstein to contribute to a survey on a controversial matter, possibly related to the political persecution of the Heidelberg mathematician Emil J. Gumbel. He refused to participate because his ability to remain in Berlin depended on his "not standing out politically on a personal level." Moreover, he did not think any Jew should be involved in this initiative. This refusal was an indication that the turbulent political atmosphere, which, in late 1923, had led to death threats against Einstein and his seeking temporary refuge in the Netherlands,[36] was still influencing the extent to which he was willing to be politically engaged and also evinces his perception of the continued uncertain plight of German Jewry (Doc. 6). Six months later, he was reluctant to publicly support the German pacifist author Heinrich Wandt, who had been sentenced to six years in prison for alleged military treason, because he did not have sufficient information on the case (Doc. 124).

In spite of his expressed desire to stay out of the public eye, this period actually saw an intensification of Einstein's public involvement. The most notable indication is the rise in the number of politically motivated appeals that he supported: from March 1926 to April 1927, Einstein cosigned eleven appeals.

Many of the judiciary of the Weimar Republic had been judges and state prosecutors for decades during the Wilhelmine period and remained loyal to the Reich, rather than to the new republic. They persisted in favoring right-wing offenders and harsh prosecution of the accused from the left.[37] The crackdown on left-wing political and cultural activities during the period of this volume preoccupied Einstein. In October 1925 he added his signature to an appeal against the State Court for the Protection of the Republic for its increasing prosecution of several left-wing artists and writers under the "Law for the Protection of the Republic," which had been enacted in July 1922 (Doc. 83).

In March 1926, he signed an appeal that demanded the expropriation of the former German royal houses without compensation (Doc. 209). He had initially objected to the phrase that the palaces and parks had been "the pleasure gardens for the mistresses of princes," which the final version of the appeal omitted. In this noteworthy instance of Einstein's advocating a nonconfrontational approach in the political arena, he opposed "the pointless exacerbation of antagonisms, which are, in themselves, necessary and productive. Excessive ranting does not win trust and educate people to become citizens of the Republic" (Doc. 206).

In June 1926, he endorsed an appeal to benefit destitute children of political prisoners (Doc. 301) and also protested the imminent ban of the pacifist film *Battleship Potemkin* by the "reactionary government of Württemberg" (Doc. 311). In August 1926, he cosigned a protest against "the white terror" of Communist activists in Poland that also demanded radical reforms of the Polish prison system (Doc. 344). The following month, he contributed a message for the protest meeting against the "bill for the protection of minors against obscene and pulp literature," organized by the Vereinigung linksgerichteter Verleger. Even though he did believe there was "a literature that does indeed have a harmful influence on young people," he deemed "the evils that such a law would entail [to be] intolerable" (Doc. 367).

In October 1926, Einstein supported an appeal against the proposed censorship law and in favor of the abolition of the Reichsprüfstelle, the national censorship board (Doc. 380). The following month, he contributed to a publication of the Rote Hilfe in its efforts to draw attention to the "systematic persecution directed against a relief organization of German workers." He opposed the "grim injustice" caused by the hampering of mutual aid efforts, but his criticism was tempered by a rather moderate stance that "painful hardships and injustices arise in this country out of a mutual unfamiliarity and lack of understanding among the classes," thereby implicitly rejecting the view that the oppression was intentional (Doc. 416). In February 1927, he cosigned an appeal with Romain Rolland and Henri Barbusse decrying the threat to political liberties by violence in most countries, voiced opposition to "white terror," and announced the establishment of a committee "to combat the wave of fascist barbarity" (Doc. 472).

Einstein's tendency to avoid direct political confrontation was also apparent in his reaction in February 1927 to the request of Hein Herbers, editor of the pacifist *Das Andere Deutschland*, to be deposed in his appeal against his conviction for inciting Reichswehr soldiers to disobedience. Einstein agreed to be deposed yet argued that "in order to win new friends, one should not repel those not close to one's own point of view by offending so-called traditional values. One must be content with opening people's eyes" (Doc. 489).

As in previous volumes, European reconciliation, particularly among representatives of the scholarly community, remained an important issue for Einstein. His involvement in the cause of rapprochement played itself out mainly on two fronts: his ongoing participation in the League of Nations' International Committee on Intellectual Cooperation (ICIC) and his efforts to end the boycott of German scientists by their European counterparts.

In the previous two years, Einstein had grown increasingly optimistic about the positive role the League of Nations could play in intra-European rapprochement. However, by June 1925, he was disenchanted by slow progress (Doc. 2). Nevertheless, in July he participated in the sixth session of the ICIC in Geneva (see Illustration 9), but decried the disproportionate influence of the French on the committee (Docs. 35 and 58). The inauguration of the International Institute of Intellectual Cooperation was held during the seventh session of the ICIC in Paris in January 1926.[38] The festivities for the opening of the new institute, at which Einstein was a guest of honor, were a significant institutional and symbolic event for European reconciliation. In his official statement Einstein again stressed the importance of the recently signed Locarno Treaties, which demonstrated that the governments of Europe had realized that the "latent struggle of traditional state entities against each other" had to cease in order for Europe to thrive. However, this goal could not be achieved merely by treaties but had to be accompanied by a parallel "preparation of the minds" conducive to a sense of solidarity among people (Doc. 165). In his notes for a toast at the inauguration banquet, Einstein expressed his gratitude toward the French nation for its role in the establishment of the institute and pleaded for scholars and artists to be liberated from "the spell of nationalism" (Doc. 167). After his return to Berlin, he reported to Paul Painlevé, with whom he had discussed the boycott of German scientists, that the German government favored the reestablishment of ties between German and Austrian scholars and societies, and international organizations. However, they had to proceed with caution so as not to alarm the more conservative minded among German scholars (Doc. 173) At the eighth session of the ICIC, held in Geneva in July 1926, Einstein was closely involved in plans to establish an international bureau of meteorology and in efforts for "world synchronization" (Docs. 331 and 332).

Einstein's views on German and European scholars during this period were strongly influenced by his perception of their positions on the ongoing reconciliation in the scientific community, in general, and the ending of the boycott of German scholars, in particular.

In his tribute to Romain Rolland, written in August 1925, he praised those few like Rolland who "cling to the ideal of the love of mankind" and decried their being "cast out by society and persecuted as lepers." In his opinion, this was "a shameful period for Europe" (Doc. 48). However, the Locarno Treaties of October 1925 renewed his optimism in European reconciliation (Doc. 94). Summing up the year 1925, he decided that Locarno was "the best thing" that had happened. The treaties also led Einstein to the conclusion that the politicians were "more sensible" than the scholars and only underscored his disappointment with the meager role European, and in particular, German, academics had played in the reconciliation among the various national scientific communities. He also saw the treaties as proof that "traditional prejudices" had been weakened in the general public (Docs. 138 and 143).

The issue of the ongoing boycott also led to controversy within the Prussian Academy of Sciences. When Einstein penned a draft for the academy's official tribute to H. A. Lorentz on the occasion of the fiftieth anniversary of his doctorate in the fall of 1925, one of the paragraphs did not find favor with Gustav Roethe, the academy's presiding secretary. Einstein had praised Lorentz's character and his crucial role in trying to heal the wounds in Europe and foster collaboration among its scientific community (Doc. 116). Max Planck subsequently informed Einstein that the paragraph had to be dropped. Einstein claimed that he "couldn't care less" (Docs. 122 and 124).

The fact that European rapprochement had been advanced more by politicians who successfully negotiated the Locarno Treaties than by scholars evoked a bitter reaction by Einstein: the professors "would have been good as avant-garde; as reserve forces they are irrelevant" (Doc. 114). However, a year later, he sounded far more positive about the role of scholars, stating that he was pleased that the leading French and German intellectuals were beginning to place themselves in the service of rapprochement between the two nations (Doc. 431). Yet he still had to deal with harsh criticism from right-wing nationalists who vociferated that the "Berlin Zionist Professor Einstein" was the "sole 'German' scientist" participating in the meetings of the ICIC. In reaction, Einstein lamented that German academics were "still caught in petty prestige politics" and emphasized that reconciliation was "the great task of our generation" (Docs. 455 and 456).

Einstein also lent his support for the public advancement of European rapprochement. In the spring of 1926, he cosigned an appeal initiated by the Verband für europäische Verständigung. The appeal argued that in the wake of the Locarno Treaties, not only the politicians but also the people had a responsibility to foster understanding and reconciliation among the European nations (Doc. 232).

Momentous steps toward rapprochement in the European scientific community were taken during this period. Following the lifting of the ban of Germans from official involvement in the Solvay Institute, Einstein was offered an appointment to the administrative committee of the institute in April 1926. In reaction, Einstein expressed his admiration that the Belgians had overcome their resentment against the Germans in favor of reconciliation (Docs. 254 and 272). A significant development in the participation of Germans in the Conseil international de recherches took place the same month. After strong opposition among conservative German academics had been overcome, Einstein reported to Lorentz that the Prussian Academy was no longer opposed to Germany's joining the council (Doc. 255).

Einstein continued to support pacifist causes during this period. In August 1926, he backed the "International Manifesto against Compulsory Military Service," initiated by conscientious objectors. In light of the ongoing disarmament in Europe, the appeal called for a "moral disarmament" and demanded that the League of Nations endorse the abolition of conscription (Doc. 357).

The issue of capital punishment was one that occupied an important place in the public discourse of the Weimar Republic.[39] In January 1927, Einstein was asked to contribute to a full-page spread in the *Berliner Tageblatt* on the abolition of the death penalty. Perhaps surprisingly in light of Einstein's long-established public image as a committed humanist, his contribution was one of three arguing *against* the abolition of capital punishment. The issue of the death penalty was a particularly controversial one at the time and had garnered heightened attention in the press as a result of two high-profile capital cases—the train massacre at Leiferde near Hanover, and a Dresden murder case. In his brief piece, Einstein argued that "in principle, I cannot understand why society should not be allowed to weed out individuals who have proven themselves to be vermin of society."[40] He also maintained that it was wrong to argue that the death penalty had "a brutalizing effect on the survivors," as this would only occur if punishment were misunderstood "as an act of retribution instead of an expression of society's striving for perfection" (Doc. 462). In the past, Einstein had appealed for specific death sentences to be avoided or repealed.[41]

Intriguingly, merely two weeks after he had argued in favor of capital punishment in general, he cosigned an appeal for clemency for those sentenced to death for the train massacre that had, in part, led to Einstein's statement in the first place (Doc. 476). He thus continued to oppose specific instances of the death penalty. He did so, for example, in the case of the murder trial against the Italian-American anarchists Sacco and Vanzetti. In April 1927, only one day after they were sentenced to death, he cosigned a cable to President Calvin Coolidge with Romain Rolland and Henri Barbusse, demanding that they be released from imprisonment (Doc. 511).

In February 1927, he contributed a short piece to a commemorative issue to celebrate ten years of the Prague Urania, an institution dedicated to the popularization of science, which was directed against the elitism of intellectuals.[42] In his statement, Einstein attacked the "certain arrogance of intellectually productive people" and criticized their lack of awareness that their opportunity to be creative intellectually was due to "the freedom enabled by the labor of others" (Doc. 491).

### VII.   Of Waves and Particles: The Emil Rupp Affair

Sometime around March 1926, the astronomer Walter Grotrian drew Einstein's attention to a recent paper by Emil Rupp (*Rupp 1926a*) on the interference of light emitted by canal rays. Leo Szilard obtained a copy of the paper for Einstein (Doc. 221), and two days later Einstein submitted a "Proposal" to the Prussian Academy for a similar experiment, one that would decide whether excited atoms emit light instantaneously (in quanta), or in a finite time (in waves). Einstein expected quantum emission to be confirmed by experiment (Doc. 223).

This was not the first time that Einstein had turned to canal rays for settling this issue. In 1922 he had suggested that, if the rays were allowed to pass through a dispersive medium, an observed deflection would corroborate their wave character, while a lack of deflection would demonstrate that they consist of quanta (*Einstein 1922f* [Vol. 7, Doc. 68]). Because no deflection was observed at the time, Einstein interpreted the result as "a refutation of the field theory of electricity." Soon thereafter, however, he had to admit that wave theory would have given the same result (Einstein to Paul Ehrenfest, 26 January 1922 [Vol. 13, Doc. 37]). The following year, Compton's experiment sparked anew his hope for an *experimentum crucis* between wave and corpuscular theory (*Einstein et al. 1923* [Vol. 14, Doc. 11]).

Rupp's results now seemed to open a new opportunity. In a paper based on his doctoral dissertation, he reported that he had investigated the light produced by canal rays with a Michelson interferometer and determined the maximum coherence length of the light. For the case of $H_\beta$ canal rays this turned out to be 15.2 cm. Eduard Rüchardt doubted that Rupp's experimental setup could have yielded the purported results (*Rüchardt 1926*). Robert d'Escourt Atkinson complained about the absence of details on the manner in which Rupp had compensated for the Doppler shift caused by both the motion of light-emitting particles in the beam and their random thermal motion (*Atkinson 1926*). Philipp Lenard, director of the Heidelberg institute where Rupp carried out the experiments, later remarked that Rupp made his observations of the interference fringes with the naked eye, but that he himself could not see anything because of their low intensity (*Dongen 2007a*, p. 112).

In Doc. 223 (*Einstein 1926p*), Einstein proposed an experiment in which the light emitted by the canal rays would be directed at a wired grid and analyzed by a Michelson interferometer using Rupp's techniques. Einstein expected that, if the wave theory was right, this should result in an "intermittent wave train." If the interference pattern was not influenced by the presence of the grid, then, Einstein argued, it would follow that the interference patterns of light had nothing to do with the atom emitting it. This, he believed, would speak against the wave theory of light. Given a grid spacing of 1/10 mm, Einstein determined that the wave trains produced by hydrogen canal rays should be 6 cm long; that is, within reach given the precision obtained by Rupp for this type of experiment.

Einstein's "Proposal" elicited criticism. Georg Joos, skeptical of Rupp's results, intended to show that the experiment would lead to the same result regardless of whether light is emitted in waves or in quanta (Doc. 266). Paul Ehrenfest shared this opinion (Doc. 248). Gustav Mie offered a special relativistic consideration that predicted needle radiation, that is, quantum emission (Doc. 268).

Einstein approached Rupp on 20 March 1926, before these critical remarks had reached him (Doc. 231), and asked him to perform his proposed experiment. Rupp accepted and said that he had been planning a similar experiment (Doc. 233). Einstein was pleased. He sketched several setups and proposed to publish the eventual results jointly (Doc. 240).

The present volume contains as full text thirty letters exchanged between Einstein and Rupp, at the time *Privatdozent* in physics at the University of Heidelberg (see Illustration 26), whose daring papers eventually turned out to be based on non-reproducible experimental results (*Dongen 2007a*). Rupp repeatedly wrote that he had experimentally proved what Einstein wanted to see, and Einstein responded with detailed criticism, pointing out how the experimental setup ought to be modified before the results could be trusted.

Rupp carried out the experiments suggested by Einstein between late April and early November 1926. Throughout, they kept in written contact. Rupp used two types of experimental setup: one contained a wire grid, while a second, also proposed by Einstein, contained a lens in place of the grid. In both cases a Michelson-type interferometer was attached to the detector. In the second type, the Doppler shift resulting from the motion of light sources along the beam was compensated by turning one of the mirrors of the interferometer by a small angle (Doc. 277). Since Rupp had observed interference, Einstein now expected the wave theory to prevail, rather than the corpuscular theory, as he had hoped in his "Proposal" (Doc. 223).

At the end of May 1926, Einstein was satisfied with the results of the grid experiments, so much so that he thought the turned-mirror experiment would add nothing to their corroboration (Doc. 299). Rupp, however, performed the experiment with various settings and observed interference (Doc. 306).

Einstein reported to Lorentz and Mie that everything confirmed the wave theory (Docs. 292 and 310). He had already formulated his arguments in Doc. 278 (*Einstein 1926v*), and discussed his paper, and Rupp's findings, at the Wednesday colloquium of 7 July 1926 (Doc. 315). Rupp's paper not being ready, Einstein only presented his own paper to the Prussian Academy the next day, but requested that its publication be delayed until Rupp could complete his own. He was "happy with the explanation of the elementary processes of optics" (Doc. 361). On 21 October, Einstein submitted Rupp's paper as well, after having corrected some of its theoretical statements (Doc. 387, *Einstein 1926w*).

Since Rupp's original manuscript is not extant, one can reconstruct Einstein's objections to the original only from his letters (Docs. 384, 389, and 395). He explained to Rupp that the build-up of the "interference field" must be distinguished from the emission of energy. The experiments do not say anything about whether the production of an elementary interference field takes place before the excited atom emits quanta or not, only that this emission time is finite and comparable to the classical damping time. Einstein was convinced that atoms radiate the interference field in an excited state during the damping time, but "we cannot conclude that the atom goes from excited state to the non-excited state *gradually*" (Doc. 384).

Rupp had apparently concluded in his manuscript that the experiments proved that the energy emission of excited atoms takes place only in waves, while Einstein was of the opinion that the "wave–quantum duality is increasingly coming to a head: energetically directed eventful process; geometrically undirected uneventful process. When will one really understand all of this from a simple foundation so that one can gain some sense of its necessity?" he wrote to Mie (Doc. 292).

The Einstein–Rupp correspondence is indeed peculiar. Whenever Einstein criticized Rupp's experiments, within days Rupp would repeat them and obtain results entirely in accord with Einstein's theoretical expectations. On many occasions, Rupp claimed to have "anticipated" Einstein's suggestions (Doc. 313) and to have "unconsciously" adjusted his instrument's settings so as to compensate for disturbances (Doc. 354).

After the two papers *Einstein 1926v* and *Rupp 1926b* were published in November 1926, Lenard remarked to Wilhelm Wien that he had not heard or seen anything of these experiments, and neither had anyone else in his institute. "Mr. Einstein may be satisfied with this work," he wrote, "I would not attach much value to it." Moreover, Lenard added, Rupp had dismantled the experimental setup before leaving for Göttingen, "if it had ever been in a proper state at all" (Lenard to Wien, 9 January 1927, cited by *Dongen 2007a*, p. 113). Yet time and again Einstein stated that he was satisfied with the results. Whenever Rupp explained away criticism leveled against the work by James Franck (Doc. 354), Georg Joos (Doc. 391), or Robert d'Escourt Atkinson (Doc. 354), Einstein stood by him (Docs. 361, 395).

Over the following years, evidence accumulated that there was something significantly wrong with Rupp's data, but Einstein did not comment. Later repetitions of the experiments lead his contemporaries to conclude that Rupp had committed fraud (*Dongen 2007a* and *2007b*).

## VIII.   Zionism and the Hebrew University

Although political and organizational anti-Semitism had become less pronounced in Germany than earlier, social and cultural prejudices against Jews became more salient during the period covered by the present volume. Anti-Jewish attitudes gained greater acceptance among the upper echelons of educated Germans. Students continued to be particularly susceptible to anti-Semitism, and anti-Jewish sentiment in the churches and among the Social Democrats was more pronounced as well. The number of violent acts against individual Jews decreased, but Jewish buildings and cemeteries continued to be desecrated.[43]

For much of this period, the Zionist movement was in crisis. At the 14th Zionist Congress in Vienna in August 1925, a challenge against the leadership of Chaim Weizmann, president of the Zionist Organisation, and his General Zionist faction emerged from the right-wing, revisionist, and religious Zionist parties, who increasingly opposed his Palestine settlement policy, his collaboration with left-wing Zionists, his role in the planned expansion of the Jewish Agency, and his policies toward the Arabs in Palestine.[44] This opposition to Weizmann was emboldened by the massive wave of middle-class immigrants to Palestine, particularly from Poland, which eventually became known as the Fourth Aliyah. The Zionist Congress ended in an impasse, and, as a result, a new Zionist Executive was not elected, thus rendering the organization without an effective leadership.[45] The majority of German Zionists supported Weizmann and his General Zionist faction.

As a counter-movement to the revisionists, the pacifist Brith Shalom (Covenant of Peace), founded in Palestine in 1925, advocated equal rights for Arabs and Jews in a bi-national state. Many of its key figures were German-speaking Jews, such as Arthur Ruppin, Robert Weltsch, Hans Kohn, and Hugo Bergmann, some of whom had close ties to Einstein.[46] Before the Vienna Congress, in June 1925, Weizmann had decided to resign from both his presidency of the Zionist Organisation and from the Hebrew University's board of governors (Doc. 14) owing to the growing internal political opposition to his policies. Einstein supported the former but not the latter action (Doc. 15).

The Hebrew University of Jerusalem on Mt. Scopus was inaugurated with great fanfare on 25 April 1925. Away in South America at the time, Einstein was the guest of honor at a celebration held in Buenos Aires on the occasion of the event.[47] This watershed moment in the nascent university's development led in July 1925 to the decision of the Zionist Executive, the supreme administrative body of the Zionist Organisation in London, to transfer governance of the university to its new board of governors. As the executive wanted the most prominent intellectuals among diaspora Jewry to exert the main influence on the board, it proposed the cooptation of additional scholars. To this end, a meeting of the European members of the board was planned for Munich in September (Doc. 24). Einstein was apprehensive that the university's local administration in Jerusalem would oppose this move. In advance of the meeting, he drafted a letter to Judah L. Magnes, chairman of the board, in which he asserted that it would be "highly dangerous" if the selection of scholars and the determination of budgets were decided in Jerusalem. In his opinion, in light of the sparse settlement of the country, the "intellectual abilities" of the scholars in Palestine were inadequate for these initial tasks, and he warned that nepotism would likely be exercised (Doc. 61). However, it is unclear whether this letter was actually sent to Magnes.

The planned gathering was held 23–24 September in Munich. In attendance were the European members of the board, as well as Magnes from Palestine and Julian W. Mack from the United States. Provisional regulations for the administration of the university until the framing of its constitution were discussed and decided upon. The provisional roles of the university's governing and administrative bodies were defined. The board elected both a presiding committee (Präsidium) and an executive committee (Exekutive), consisting of two members each. Einstein and Weizmann were elected presidents of the board; Magnes and Norman Bentwich were elected members of the executive committee, as chancellor and vice-chancellor, respectively. The meeting also decided on the expansion of the board: nineteen new members were chosen, and three more members were to be selected by the Americans. The new members were European and American schol-
ars and financial backers from the United States.[48]

Weizmann, who was not present at the Munich meeting, deemed some of the
new members incompetent. He was also unhappy with the new, British university
titles conferred on Magnes and Bentwich, unsuitable, to his mind, for the adminis-
tration of the university (Abs. 124).[49] Einstein concurred and wrote to Weizmann
that he would "never forget what an embarrassing impression" Magnes's bestow-
ing the title of chancellor upon himself had made on him. Magnes needed to be told
that he was the executive organ of the board. He confirmed the accuracy of the min-
utes of the meeting and objected to their being presented to Magnes prior to their
distribution to the other members of the board. Einstein warned that should Magnes
continue to behave in this manner, he would have to be removed, regardless of
whether he had backing from the American funders of the university (Doc. 117).
Thus, the opening shot was fired in what would evolve into a major conflict be-
tween Einstein and Magnes over the minutes of the Munich meeting.[50]

In late December 1925, Einstein, as president of the board, protested to Magnes
against his sending out a second version of the minutes. He was disturbed by
Magnes's alternative version, whose revisions mainly dealt with the principal res-
olutions adopted by the board, and urged Magnes to withdraw the "invalid min-
utes" (Doc. 142). The version of the minutes sanctioned by Einstein was prepared
by Leo Kohn, chairman of the Zionist Organisation's university committee in Lon-
don (see Illustration 7). The second version was produced by Magnes himself. The
most significant differences related to the location of the central office of the pre-
siding committee, the definition of the role of the presiding committee, the charac-
ter and authority of the proposed academic council, the identity of the secretary of
the board, and the university's budget.[51] The tug-of-war over the minutes boiled
down to two decisive interrelated questions. First, did the main locus of control
over the governance of the university lie with the Zionist Organisation in London
and the diaspora intellectuals on the board of governors, or with the local executive
of the university in Jerusalem? Second, were the decisions taken by the board in
Munich binding, or were they merely proposals that needed to be approved by the
executive in Jerusalem?

Einstein informed Felix M. Warburg, the university's most influential American
backer, of the outcome of the Munich meeting. He stated that "nothing would be
more disastrous than to transfer the top-level governance of the university to Pal-
estine," which lacked scholars with "the essential experience and intellectual cali-
ber to build up and govern a university worthy of the entirety of diaspora Jewry."

The intellectual governance of the university had to remain outside of Palestine, as that was where Jewry's intellectual center of gravity lay (Doc. 145).

In March 1926, Einstein expressed his astonishment at Magnes's refusal to retract his version of the minutes. He had concluded that it was "futile" to continue negotiating with Magnes, and he intended to forward the matter to the board for its decision (Doc. 214). After learning of Weizmann's intention to resign from the Zionist Executive, Einstein expressed the hope that he would dedicate himself fully to the university, thus rendering Magnes superfluous (Docs. 213 and 250).

In May, after visiting the university in Jerusalem, Weizmann sent Einstein a highly critical report in which he detailed his unfavorable impressions of Magnes, who was "dilettantishly bungling about" in the Institute of Jewish Studies (see Illustration 6). He did not see any way to replace Magnes but was convinced that his "autocracy must be broken" (Doc. 281). In reply, Einstein proposed that "a person with a scientific mind and managerial and psychological predisposition" should function as the permanent representative of the board of governors (Doc. 285). In early July, Einstein wrote to the board calling for Magnes's removal. He could not continue to be involved in university matters until Magnes's resignation had been implemented. He stressed that his letter should be treated as confidential to avoid public harm to the university, but then added a note in the margin of the letter that the demand for Magnes's removal was "retracted" (Doc. 318). In his cover letter to Weizmann, Einstein clarified that he would adhere to his intention not to be involved in university matters until Magnes was dismissed (Doc. 319). Weizmann expressed great dismay and implored Einstein to tell him how to prevent such an action (Doc. 320). By the time of his arrival in Bern on his way to Geneva, Einstein had resolved to follow through with his demand for Magnes's removal (Doc. 326). Thus, merely a little over a year following the inauguration of the Hebrew University, Einstein had de facto resigned from active participation in its governance.

There followed a four-month hiatus during which no letters by Einstein on this matter are extant. In November 1926, he informed Weizmann that he could not accept reelection as chairman of the university's academic council because of "the current circumstances" (Doc. 409). No substantial correspondence followed until May 1927, except in February 1927, when numerous newspapers reported that Einstein intended to resign from the board. But Einstein denied the reports (Abs. 760). The correspondence on the Hebrew University in this volume ends with Einstein's informing Leo Kohn that his involvement was "precluded under the present circumstances." However, he would not resign formally so as to not harm the institution (Doc. 527).

Einstein continued to endorse some Zionist enterprises. In early March 1926, he attended a rally in Berlin in behalf of the German branch of the Keren Hayesod, the Palestine Foundation Fund. In a statement for the occasion, he made it clear that activity in support of Jewish settlement in Palestine did not render Jews any less German. For him, the "distinction is not between Jew and German, but rather between upstanding and spineless" (Doc. 208).

Einstein's most important contact with the German Zionist movement continued to be Kurt Blumenfeld, president of the Zionistische Vereinigung für Deutschland.[52] In response to Blumenfeld's request that Einstein support a Zionist campaign in March 1926, he stated that he had high esteem for the educational achievements of Zionism but did not know the movement well enough. He therefore asked to be sent the Zionist Organisation's annual report, information on income from the various countries, and on the use of these funds in Palestine and elsewhere, thereby indicating that he would not automatically endorse the planned campaign (Docs. 225 and 227). Half a year later, Blumenfeld expressed concern about numerous reports of Einstein's "changed position relative to Zionism." Einstein's statements were allegedly not only critical, but also derogatory. Blumenfeld acknowledged that not all of Einstein's experiences with the Zionist movement had been pleasant ones, yet he did not believe that Einstein would be decisively swayed by such interludes. He asked him to clarify his stance toward the Zionist movement, yet no reply is extant (Doc. 385). We do not have documentation that would substantiate a transformation of Einstein's position toward Zionism during this period. But it is definitely possible that his alleged dissatisfaction was related to his disenchantment with developments at the Hebrew University, by far the most important Zionist enterprise that Einstein had supported.

Nevertheless, in late 1926, Einstein backed a very important Zionist enterprise when he attended the inaugural assembly of the Deutsches Komitee "Pro Palästina" in Berlin and was elected a member of its honorary board. The committee, originally founded in 1918 and reestablished on Blumenfeld's initiative, aimed to educate the German public about Jewish settlement in Palestine and foster relations between Germany and Palestine. Its members included high-ranking officials of the German government, politicians, diplomats, and Jewish and non-Jewish intellectuals.[53]

This volume also presents two brief writings by Einstein that demonstrate the deepening reception of relativity in various sectors of Jewish culture. In July 1926, he wrote a foreword for the Hebrew translation of his exposition on special and general relativity (*Einstein 1917a*), being "especially pleased with its publication in the language of our forefathers" (Doc. 317). Later in the year, Einstein was also happy with the publication of a book on relativity in Yiddish (Doc. 423).

## IX.  The New Quantum Mechanics

Throughout his scientific life, Einstein was engaged in studying and deepening the understanding of the quantum nature of heat, light, and matter. Over a period of twenty-five years he had contributed numerous papers to research on quantum theory. In 1925, he quickly recognized the great importance of the new quantum theories; and he was equally quick in realizing the conceptual difficulties involved. As this volume shows, from the beginning Einstein preferred wave mechanics over matrix mechanics, and he emphasized the important fact that Schrödinger's wave function is defined on configuration space rather than on spacetime (Docs. 304, 307, 310, 353, 362).[54]

On 15 July 1925, Einstein received a letter in which Max Born announced: "Heisenberg's new work, soon to be published, looks very mysterious, but is certainly true and profound" (Doc. 23). Werner Heisenberg's paper was published on 18 September 1925. It was followed ten days later by a paper coauthored by Born and Pascual Jordan, and then by a third, joint paper by Born, Heisenberg, and Jordan in February 1926 (see Illustrations 22 and 24).

In his first paper, *Heisenberg 1925*, Heisenberg reinterpreted the classical equations of motion for position and momentum as expressing relations between arrays of numbers (hence the term "Umdeutung" in the title of the paper); Jordan and Born recognized these arrays as matrices with infinitely many rows and columns subject to a noncommutative multiplication law in *Jordan and Born 1925*; and in their joint *Born, Heisenberg, and Jordan 1926* paper, the authors extended the procedure to systems with arbitrarily many degrees of freedom. These three papers laid the foundations of the new matrix mechanics.

Einstein appreciated the ingenuity of the new approach but was dissatisfied. Two days after the publication of Heisenberg's first paper he wrote to Ehrenfest: "Heisenberg laid a big quantum egg. In Göttingen they believe in it (I do not)" (Doc. 114). He began a lively correspondence with Heisenberg and Jordan. Unfortunately, Einstein's letters to both of them are not extant, with the exception of one postcard to Jordan (Doc. 212).

Nevertheless, we can reconstruct Einstein's comments and criticism from the long letters that both Heisenberg and Jordan wrote back to him, from which it becomes clear that Einstein's letters, too, must have been extensive. The main part of the correspondence with Einstein took place while Born, Jordan, and Heisenberg were working on the definitive form of their new quantum mechanics. Indeed, in his first letter to Einstein on 27 October 1925 (Doc. 98), Jordan sent draft notes of what would eventually become the landmark paper *Born, Heisenberg, and Jordan 1926*.

One major topic in the correspondence concerns a peculiar result following from the treatment of the harmonic oscillator, which serves as a model for black-body radiation in the new formalism. It was required that in the ground state a harmonic oscillator does not emit any radiation. From this it followed, within matrix mechanics, that a harmonic oscillator has a zero-point energy of $\frac{1}{2}h\nu$. Einstein must have criticized this result, because both Heisenberg and Jordan agreed that it was problematic. Heisenberg argued that the zero-point energy was "initially purely formal. In any event, its physical meaning has not yet been completely clarified" (Doc. 112). Jordan wrote that it is "actually just a formal quantity, without any direct physical meaning" (Doc. 132). While Einstein's reply to them is unknown, it probably was similar to his comments to Ehrenfest two months later: "There can't be any zero-point energy in cavity radiation. I deem the argument by Heisenberg, Born, and Jordan (fluctuations) as faulty, if only because the probability of large fluctuations... certainly does not come out correctly that way" (Doc. 194). The calculation and interpretation of the zero-point energy of a harmonic oscillator played a major role in almost all other exchanges as well (Docs. 119, 198, 199, 212, 247, 524). Other important topics were the interpretation of the canonical commutation relations (Docs. 194, 198), electron spin and thus half-integer quantum numbers (Docs. 112, 132, 199), and the classical limit of matrix mechanics.

Einstein's reception of matrix mechanics is expressed in several letters to friends and colleagues. In his Christmas greetings to Besso he wrote: "The most interesting thing that theory has produced recently is the Heisenberg-Born-Jordan theory of quantum states. Such a magic formula, in which infinite determinants (matrices) take the place of Cartesian coordinates. Highly ingenious, and sufficiently protected from refutation by great complexity" (Doc. 138). To Hedwig Born, who had just returned from accompanying her husband on a lecture tour to the United States, Einstein expressed himself in a more positive vein: "The Heisenberg-Born ideas are keeping everyone breathless, the reflection and thinking of all theoretically interested people. In place of dull resignation, a unique tension has taken hold in us sluggish people" (Doc. 215). In letters to Lorentz and Zangger (Doc. 243) he was rather more outspoken. He wrote to Lorentz: "I have been quite occupied with Heisenberg-Born. With all due admiration of the intellect behind these papers, my instinct still balks at this kind of conception" (Doc. 218). Lorentz felt the same (Doc. 220).

The flip side of the coin of what is today called quantum mechanics concerns the formulation of wave mechanics by Erwin Schrödinger (see Illustration 21). In the summer of 1924, Louis de Broglie had suggested in his dissertation that matter may possess wavelike properties. His adviser, Paul Langevin, sent a copy of the dissertation to Einstein (see Vol. 14, Introduction, p. lxxxiv), who discerned its revolutionary nature and responded on 16 December 1924: "The paper by De Broglie greatly impressed me. He has lifted one corner of the great veil" (Vol. 14, Doc. 398). On the same day he also wrote to Lorentz: "I think this is the first weak ray to cast light on this worst of all our physical puzzles (i.e., the Bohr-Sommerfeld quantum rule)" (Vol. 14, Doc. 399). Einstein drew on De Broglie's ideas in his papers on the quantum theory of the ideal gas, especially in *Einstein 1925f*, where he wrote: "One then sees that to such a gas a scalar wave field can be assigned,… it seems as if a wave field were associated with every process of motion" (Vol. 14, Doc. 385).

It now appears that Schrödinger's wave mechanics was at least partly inspired by his discussions with Einstein on Bose-Einstein statistics and the link that Einstein had forged between quantum statistics and De Broglie's idea of matter waves. Their remarkably rich exchanges on the topic began on 26 September 1925. Einstein wrote how much he appreciated Schrödinger's recent publications on quantum statistics but then spent most of the letter criticizing Max Planck's recent paper on the same topic and asking for Schrödinger's opinion (Doc. 80). The ensuing letters (Docs. 101, 103, 107, 108, 123, 174) center on the various ways of deriving and interpreting quantum statistics, as well as on the observation that the particles derived by it are "squatting together," as Schrödinger puts it—the phenomenon now known as Bose-Einstein condensation (Doc. 101). Schrödinger saw himself as only working out Einstein's original idea (in Doc. 80) and suggested a joint publication. Einstein answered: "I just don't know whether I should count as a coauthor since after all you did all the work; I would feel like an 'exploiter,' as the socialists like to put it so beautifully" (Doc. 108). To this Schrödinger replied: "The idea of regarding you as an 'exploiter' would not have occurred to me even in jest. To continue with the sociological work analogy, one could rather say: When a sovereign builds, carters have plenty to do" (Doc. 123), a reference to Goethe and Schiller's *Xenien*, in which the two poets refer to Immanuel Kant as the king and to his interpreters as the carters. How Einstein could have resisted such an invitation we shall never know. He presented Schrödinger's sole-author paper to the Prussian Academy on 7 January 1926, he himself having supplied an abstract (Doc. 153).

Schrödinger was getting closer to becoming royalty himself. Three weeks before Einstein presented *Schrödinger 1926a* to the Academy, Schrödinger submitted his paper "On Einstein's Gas Theory" to the *Physikalische Zeitschrift* (*Schrödinger 1926b*). It was soon followed by his first paper on wave mechanics, "Quantization as an Eigenvalue Problem. First Communication," published in *Annalen der Physik* on 13 March 1926 (*Schrödinger 1926c*), which Einstein must have seen for the first time during the second week of April. In a letter to Zangger early that month he had complained: "I like De Broglie's idea, but so far nothing can be done with it" (Doc. 243). But nine days later, he praised Schrödinger in a letter to Ehrenfest: "The Born-Heisenberg thing is probably not correct, after all. It does not seem possible to make the correspondence between a matrix function and a normal one unique... In contrast, Schrödinger created a very different and highly clever theory of quantum states by allowing De Broglie's waves to act in phase space... Not such an infernal machine but rather a clear idea and 'unavoidable' in its application" (Doc. 253). Similarly, he expressed excitement in his letter of the same day to Lorentz, recommending Schrödinger as a speaker at the upcoming 1927 Solvay conference instead of himself: Schrödinger "has a theory of quantum states in press, a truly brilliant implementation of De Broglie's idea" (Doc. 254). Lorentz responded by clarifying that he envisaged two quantum lectures at the Solvay conference, one of which would address fundamental quantum dynamics, to be given by Heisenberg or Schrödinger, and a second on quantum statistics. Lorentz hoped that Einstein would take it upon himself to speak on the latter (Doc. 269).

Einstein agreed and again praised Schrödinger, whose "version of the quantum rule impresses me much; this seems to me to be a piece of the truth, even though the meaning of waves in an $n$-dimensional $q$-space remains so much in the dark." For the first time Einstein pointed out what he considered to be the crucial problem with Schrödinger's wave function, namely, that it is not defined on spacetime, but on configuration space, where $n$ is the number of particles in the system described by the wave function (Doc. 272). He also raised this issue in letters to Besso, Epstein, Sommerfeld, and Ehrenfest, and complained to the latter about the unintelligibility of Paul Dirac's recent work as well (Docs. 271, 304, 353, 356, 362).

Einstein began the correspondence with Schrödinger on wave mechanics shortly afterward. On 16 April 1926 he wrote: "Mr. Planck has shown me your theory with justified enthusiasm, which I then also studied with the greatest interest." He immediately started to tackle the details of the theory, but in the margin he added: "The idea of your work shows real brilliance" (Doc. 256). Schrödinger replied warmly: "Approval by you and Planck is more valuable to me than approval by half the world... Besides, this whole matter would not have been created now or ever (I mean, by me) if your second gas degeneracy paper had not shoved the importance of De Broglie's ideas in front of my nose" (Doc. 264).

Five days later, on 28 April 1926, Heisenberg delivered a lecture in the Berlin Physics Colloquium. Afterward, Einstein invited him to his home. In Heisenberg's much later recollections (*Heisenberg 1969*), he credited Einstein with having laid the foundation, during their conversation, to his own discovery of the uncertainty relations in February 1927, almost a year later.

Having looked at how Einstein reacted to, and indeed partly inspired, the discovery of both matrix and wave mechanics, we shall now turn to his reaction to the emerging dominant probabilistic interpretation of the new quantum mechanics.

In his letter of 18 February 1926, Heisenberg had already advocated a statistical interpretation (Doc. 198). But Einstein's first explicit reaction is found in reply to Gustav Mie's argument in April that causality had to be abandoned altogether and replaced with statistics (Doc. 268). For Einstein, this was too big a step: "Doing away with strict causality need not be final, for Heisenberg's theory does not claim to be a complete theory at all, but is rather just a mathematical version of the correspondence principle" (Doc. 292). This is the first time in Einstein's correspondence that the problem of jettisoning causality was raised in direct connection with the new quantum mechanics, although it had arisen previously in connection with the Bohr-Kramers-Slater theory and the Compton effect (see Vol. 14, Docs. 240, 256, 259).

A few months later, Born supplied his own interpretation of Schrödinger's wave function as giving the probability amplitude of finding quantum particles in a particular state (*Born 1926a, 1926b*). For Born, this interpretation built on an idea expressed by Einstein five years earlier (*Einstein 1922a* [Vol. 7, Doc. 68]), as he emphasized both in the papers and in a letter to Einstein: "About myself I can report that in physics I am quite satisfied, because my thought of conceiving Schrödinger's wave field as a 'ghost field' in your sense is increasingly proving its worth." He elaborated on this probability field: "Pauli and Jordan have made fine advances in this direction. Naturally, the probability field does not move in normal space, but in phase (or configuration) space." He then concluded: "Schrödinger's accomplishment reduces to something purely mathematical; his physics is quite meager" (Doc. 422).

Einstein responded with what has by now become one of his best known and most quoted letters, albeit in varying translations: "Quantum mechanics is very worthy of respect. But an inner voice tells me that it is not the genuine article after all. The theory delivers much, but does not really bring us any closer to the secret of the Old One. I, at any rate, am convinced that *He* does not play dice" (Doc. 426).

Despite his opposition to Born's interpretation of Schrödinger's wave mechanics, the theory itself continued to preoccupy Einstein. In February 1927, writing to Lorentz, he noted: "Quantum theory has been completely Schrödinger-ized and has much practical success as a result. But surely it cannot be a description of the real process. It is a mystery" (Doc. 479). He wrote to Zangger in a similar vein a month later (Doc. 507).

In the present volume, the only letter by Niels Bohr dates from 13 April 1927 (Doc. 513). At Heisenberg's request, Bohr sent Einstein a copy of the proofs of *Heisenberg 1927*, in which Heisenberg introduced the uncertainty relations. Bohr's accompanying letter contains a detailed discussion of the uncertainty relations as applied to a monochromatic plane light wave. He praised Heisenberg's paper as signifying "a most meaningful contribution," and discussed the connection between quantum and classical physics in light of the uncertainty relations. We may detect here the seeds of Bohr's version of the Copenhagen interpretation and the beginning of the enduring Einstein–Bohr dialogue on the interpretation and validity of quantum mechanics.

As far as we know, Einstein did not answer Bohr's letter. But Born's probabilistic interpretation of Schrödinger's wave function and Heisenberg's uncertainty relations were no doubt on his mind when, three weeks later, he completed a manuscript titled "Does Schrödinger's Wave Mechanics Completely Determine the Motion of a System, or Only Statistically?" (Doc. 516). The paper contains many novelties.

First, Einstein explores not only the properties of Schrödinger's equation, but also takes up Schrödinger's idea of using Riemannian geometry to better understand the structure of the configuration space on which the wave function $\psi$ is defined. He defines a Riemannian metric on configuration space and calls the second covariant derivative of $\psi$ (with respect to the derivative operator compatible with the metric) the "tensor of $\psi$-curvature"; the twice-contracted "tensor of $\psi$-curvature" he calls the "scalar of $\psi$-curvature."

Second, Einstein spells out a notion of what it would mean to complement the description of a physical system described by the wave function in such a way that the description becomes "complete." He assumes that the configuration space is $n$-dimensional, that is, he assumes that $n$ particles move in one spatial dimension. Their description would be complete, Einstein argues, if one could associate $n$ directions to the $n$ dimensions of configuration space and if the energy of the total system could be written as a sum of $n$ terms uniquely corresponding to the $n$ directions in configuration space. For one could then associate to each of these terms a corresponding velocity, and the sum of all these velocity vectors would be the velocity vector of the total system in configuration space. This latter vector would then be completely described.

Third, Einstein implicitly defines a variational problem in terms of the scalar of $\psi$-curvature, and introduces Lagrange multipliers $\lambda$, which result in principal directions in configuration space, to solve the problem thus posed. It has been argued that by introducing principal directions of configuration space, Einstein effectively introduced what would today be called hidden variables. However, a distinction may be made between dynamical and nondynamical hidden variables;

Einstein's theory would belong to the latter category.[55]

On the same day in May 1927 that Einstein presented this paper to the Prussian Academy, he wrote to Ehrenfest that he had shown that "solutions [to the Schrödinger equation] can be uniquely assigned to particular motions, which makes any statistical interpretation unnecessary" (Doc. 517). A few days later he also reported the same to Born (Doc. 520) and thereby ignited Heisenberg's "burning interest" (Doc. 524). Upon receiving the proofs, Einstein composed an addendum that suggests that his outlook on wave mechanics, and his own modification to it, had changed fundamentally within the two weeks since presenting the paper to the Academy. The addendum is concerned with an issue that Einstein had already puzzled about in his initial evaluation of both wave mechanics and matrix mechanics, an issue that resonates with questions raised almost a decade later in *Einstein, Podolsky, and Rosen 1935*.

In his first letter of 16 April 1926 to Schrödinger (Doc. 256), Einstein considered the question whether Schrödinger's equation allowed the description of what he would ten days later call the "requirement of system additivity" (Doc. 267). Einstein demanded that, if $E_1$ is a possible energy eigenvalue of a first system and $E_2$ a possible energy-eigenvalue of a second system, and if the two systems are not coupled to one another, then $E_1 + E_2 = E$ should be a possible energy eigenvalue of the total system. Believing that Schrödinger's equation lacked this property, Einstein had suggested an allegedly better one and had argued that this equation satisfied the requirement of system additivity. It turned out, however, that Schrödinger's paper had in fact contained precisely the equation suggested by Einstein as an alternative. Six days later, Einstein admitted his error in a postcard (Doc. 261). But Schrödinger did not receive the card before replying to the previous letter and pointing out that his equation was exactly the one that Einstein had suggested (Doc. 264). Nevertheless, Schrödinger was rather happy about Einstein's letter, for he saw it as reconstructing his equation from the requirement of system additivity and from a second requirement that absolute energy values are impermissible. In reply, Einstein praised Schrödinger's theory for adhering to the requirement of system-additivity, considering this a major advantage of wave mechanics over matrix mechanics (Doc. 267).

In his addendum to Doc. 516, written almost a year after the first comments on the issue, Einstein defines the requirement of system additivity in exactly the same way. He does not take back his claim that wave mechanics (and his own modification of it presented in the paper) obeys the requirement. However, he adds a second requirement, motivated by the first. He demands that if a system is indeed composed of two subsystems that do not couple, with each subsystem having a certain set of possible energy eigenvalues, then the possible motions of the total system *must* be combinations of the possible motions of the subsystems. Einstein then argues that his modified wave mechanics does not fulfill this requirement: the theory does not in general allow decomposition of the motions of a composite system into the motions of its subsystems. He justifies this claim by appeal to his tensor of $\psi$-curvature, $\psi_{\mu\nu}$. He argues that the latter does not vanish for the kind of situation discussed and that this implies that the principal directions of the composite system do not coincide with the principal directions of the subsystems if the latter are regarded as isolated systems.

Einstein evidently considered this to be a major problem. The addendum finishes with an expression of hope that it might yet be possible to overcome the issue by following through with an idea of Jakob Grommer, namely, to use lg$\psi$ instead of $\psi$ to define the principal directions. But the idea did not pan out and, only two and a half weeks after presenting the paper to the Academy, Einstein called the journal and withdrew the paper from publication.

## X.  Refrigerators and Patents

Refrigerators had already aroused Einstein's interest in December 1919 (see Vol. 9, Doc. 207) when, together with Walther Nernst, he had worked on a "funny technical thing, an ice machine." By March 1922, they were considering a patent application (Vol. 13, Doc. 67). Einstein must have been an excellent discussion partner, but it was Nernst who led negotiations with a company and worried about technicalities (Vol. 13, Abs. 63).

Leo Szilard claimed that in 1926 it was Einstein who initiated a project for developing safe household coolers after having read in a newspaper that an entire family was killed in bed by poisonous gas leaking from the pump of their refrigerator. Einstein therefore planned to develop coolers that had no moving parts and were hermetically sealed. How and why he chose Szilard as a partner is not known; perhaps he wanted to help Szilard, a man always in need of money.[56]

The first reference to their collaboration in developing refrigerators can be found in Szilard's letter of 10 September 1926 (Abs. 579). Because they had been in contact since the early twenties and were becoming closer at least since the end of 1923 (Vol. 14, Doc. 198), there was no need of communicating by letters, since they both lived in Berlin and could meet in person (Doc. 221). Szilard informed Einstein that he had submitted an application for a patent on a refrigerator (Abs. 579). An application known only by its title, "on a cooler with capillary pump," by Szilard and Einstein was filed with the German Patent Office on 13 September ([35 558]).[57] Other applications had been filed with the Patent Office in March (S73730.1/17a) and October, the latter design possessing a water steam jet refrigerant, but further details on the patents are unknown. A third application on an absorption-diffusion cooler was submitted on 26 October (patented as DE499830).

In October, Szilard approached Bamag-Meguin Co. and offered the firm three cooler inventions for purchase, financing, development, production, and sales. He entrusted his brother, Béla (Adalbert), to enter into negotiations in person ([35 560] and Abs. 624) and drafted a contract with the company (Abs. 633 and 634). After discussions with general director H. Peiser, he drew up a second draft contract (Doc. 417 and Abs. 649). Meanwhile an application on another absorption cooler was submitted on 16 December.[58]

In April 1927, Szilard was working hard on fitting a "mercury pump" in a refrigerator (Doc. 512). This pump was his own invention: an arrangement for "pouring molten metals into a mold by using electric current," for which he had submitted an application on 20 January 1926 (patented as DE 476812). He also considered the time to be ripe for entering into a written agreement with Einstein on how to share income from the refrigerators (Abs. 814).

In view of Szilard's intense activity in this cooperation, and Einstein's few letters, it is difficult to establish Einstein's contributions. The capillary pump seems to have been his idea. "I am happy […] about the capillary pump"— he wrote to Szilard on 15 September 1928 —"for which I have not yet succeeded in arousing your enthusiasm" [21 432]. But their exchanges must have been quite frequent, since Szilard noted, "soon you will be getting sick enough of ice machines" (Abs. 649). Whatever they had achieved by late spring of 1927 would be only the first stage of their future collaboration.

## XI.   A Hectic Pace of Life

Despite having tried in previous years to minimize public appearances and obligations, the rapid pace of Einstein's private and professional life seemed unrelenting.[59] In October 1926, he confided to Zangger that "so much keeps assailing me that I rarely have time for myself" (Doc. 397). His celebrity status did not help matters and irked him at times. Thus, he asked Friedrich S. Archenhold, director of the Treptow Observatory in Berlin, not to involve him publicly in a planned exhibition about the planet Mars, adding in exasperation: "Can you understand that I'm tired of appearing everywhere as a symbolic bellwether with a halo? So, leave me out of it!" (Doc. 394). He thought his work was suffering: "I myself have not managed to do anything of notable good. The muses tend to grant their favor only to the young, and that's good" (Doc. 286). And he was pleased that neither one of his sons was planning a career in science, as it was "full of futile hard work" (Doc. 257).

In November 1926, Einstein informed Ehrenfest that owing to his obligations at the League of Nations and "several industrial matters," he did not have sufficient time or energy to maintain his position as a special professor at Leyden (Doc. 420). The following month, he again referred to his undesired fame: in replying to his old flame Anna Meyer-Schmid, he remembered his youth "when one didn't yet have any gray hairs and one wasn't the afflicted big shot from whom everyone wants something" (Doc. 435). And when declining for the third time Millikan's invitation to Pasadena, he averred that he would be unable to embark on large-scale trips as he had "transformed from an animal into a plant" (Doc. 445).[60] He was aware, as he wrote to Anschütz-Kaempfe, that he was different from other scholars, who needed more relaxation "than the likes of us, who only work when they feel like it or when hit by a frenzy" (Doc. 37).

Einstein was confronted with several deaths and illnesses among family members and friends. Elsa's parents both died; his stepdaughter Ilse suffered from stomach ulcers (Doc. 42); and her husband, Rudolf Kayser, was diagnosed with a heart condition (Doc. 426). His close Dutch colleagues Willem H. Julius and Heike Kamerlingh Onnes both died. His old friend from their student days, Marcel Grossmann, was diagnosed with multiple sclerosis (see Doc. 34, note 2).

The personal and professional crisis of Einstein's close friend Michele Besso, which first surfaced in June 1925 when his work output was at "zero," is well-documented in the volume. Besso confided that he had started psychoanalysis and was hopeful about its outcome (Doc. 5). After visiting Besso in Bern a year later, Einstein concluded that his friend was "still very brilliant, but, unfortunately, not very fit for work" (Doc. 328). Five months later, Besso was in danger of losing his position at the Swiss Patent Office because of his low productivity. His son Vero asked Zangger to intervene (Docs. 401 and 405). In reply to Zangger's inquiries, Einstein described his friend as "one of the strongest, brightest minds and most sincere characters that I have met in my entire life" and praised his "astounding" knowledge in physics and technology. Yet "[h]is weak side is willpower." Asked by colleagues for assistance at the office, Besso would devote exhaustive time to aid them with their work, and, as an "overly conscientious person," his own work suffered as a result. Einstein, who could not take the initiative himself to come to Besso's assistance, as this might be perceived as being "immodest," suggested being asked to do so (Doc. 432). He subsequently wrote a more official letter to Zangger, stating that Besso's dismissal would be "a grave mistake" (Doc. 436) and an "Opinion on Michele Besso" that presented similar arguments (Doc. 451).

At times, Einstein's reaction to the turmoil in his life was fatalism. In December 1925, he wrote rather stoically that "[t]he beginning and the end of life are damned difficult, and in between things don't always run smoothly, but everything passes and is quickly forgotten" (Doc. 124). He also commented on death in a similar manner: "death is ultimately nothing but a dot at the end of a well-formed sentence" (Doc. 383). In July 1926, he wrote to Elsa during her mother's terminal illness: "You also need not expect an imminent end; that is not very likely. You must passively endure it as a necessary twist of fate and also take the good as it may be offered" (Doc. 329). In any case, the increase in the pace of life does not seem to have encouraged Einstein to greater introspection. In January 1927, Hugo Freund, a psychotherapist and Social Democratic politician, asked Einstein to participate in a study of prominent political and economic leaders who would undergo Adlerian analysis. The results of this experiment would be published in the press. Einstein replied that he "would like to remain in the darkness of not-having-been-analyzed" (Docs. 457 and 458).

In spring 1927, the municipal department of construction regulation served Einstein with an eviction notice from his attic "tower room" above his apartment; it had been deemed inappropriate for residential purposes since it was lacking in adequate sanitation (see Illustration 17). Einstein appealed to the president of police of the Schöneberg district in Berlin, arguing that he deserved special consideration as a well-known scholar and teacher at the university. Moreover, since the room was only used by himself, it would only affect him personally and no one else if the room were unhygienic (Doc. 523). This wry sense of humor continued to stand Einstein in good stead. In August 1926, having been asked to play the violin in Brahms's Sextet for the opening program of the upcoming first international congress on sexual research in Berlin, he replied that "unfortunately, neither my sexual nor my musical capabilities allow me to regard myself in a position to meet your kind request." It is unclear whether he mailed the letter (Abs. 542 and Doc. 351).

## XII. The Newton Bicentenary

At the time when the new developments in quantum mechanics were holding everyone in physics "breathless," as Einstein wrote to Hedwig Born (Doc. 215), classical physics had reason to remember one of its greatest heroes, Isaac Newton. The bicentenary of his death was to be commemorated across the world on 31 March 1927 (see Illustrations 16 and 17).

Shortly beforehand, Eddington informed Einstein that a celebration would take place in Grantham, England, where Newton had gone to school, to be attended by many leading physicists and mathematicians. He asked whether Einstein would be willing to send a "a message of good wishes" to the gathering, given his admiration for Newton (Doc. 492). Einstein obliged, but his short text did not reach the secretary of the Royal Society, James H. Jeans, in time to be read at the event (Doc. 494).

This was not the only occasion for Einstein to praise Newton in the months leading up to the bicentenary. Indeed, as he complained to Paul Feldkeller in the spring of 1927 (Doc. 496), he had been overwhelmed by similar requests. In the end, in addition to the message for the Grantham gathering, a translation of which would be printed in *Nature* (Doc. 504), Einstein wrote more expansive articles for *Die Naturwissenschaften* (Doc. 503), published in English in the *Manchester Guardian*, and for the journal *Nord und Süd* (Doc. 506), the latter being the manuscript for a radio broadcast read by Einstein on the day of the bicentenary.

In each of these three texts, Einstein presented Newton as the champion of strict causality in physics, which he identified with the ability to deduce the state of motion of a system from its immediately preceding state (Doc. 503). He stated that Newton was the first to express the causality requirement in a rigorous manner by writing the laws of physics in the form of differential equations. "What has happened since Newton in theoretical physics," Einstein wrote, "is the organic development of his ideas." He presented the introduction of the concept of a field in the nineteenth century as the next significant step in this development. Newton's differential equations found their natural continuation in the partial differential equations governing fields. It is only in quantum theory "that Newton's differential method becomes inadequate, and indeed strict causality fails us." But he expressed the hope that the "last word has not yet been said" and that "the spirit of Newton's method" would eventually be restored (Doc. 494).

This aspiration was voiced about two months before Einstein presented, and then retracted, a paper in which he tried to modify Schrödinger's wave mechanics so that, had it been successful, it would arguably have led to a recovery of strict causality as defined in his texts on Newton (see sec. VIII of this Introduction).

Intriguingly, in his message to the Grantham meeting, Einstein first wrote down and then decided to strike a phrase that he had already used in a slightly different form in a letter to Max Born, one that would later become iconic: "God does not play dice" (Doc. 426).

He concluded his paper in *Naturwissenschaften* by writing: "[W]ho would be so venturesome as to decide today the question whether causal law and differential law, these ultimate premises of Newton's treatment of nature, must definitely be abandoned?" (Doc. 503).

[1]The most notable examples of right-wing opposition to relativity took place in 1920 (see Vol. 10, Introduction, pp. xxxviii–xli, and *Wazeck 2014*).

[2]See Vol. 14, Introduction, p. lxiii.

[3]See the press release of the Bernisches Historisches Museum, "Einstein zog 500.000 Besucher an—ein Grund zum Feiern," 24 March 2015. http://www.bhm.ch/fileadmin/user_upload/documents/Medien/2015/Medienmitteilung_Einstein-Programm_D.pdf.

[4]See Einstein to Marie Winteler, with a Postscript by Pauline Einstein, 21 April 1896; Marie Winteler to Einstein, 4–25 November 1925; Marie Winteler to Einstein, 30 November 1896; (Vol. 1, Docs. 18, 29, and 30).

[5]See Einstein to Pauline Winteler, May? 1897, and 7 June 1897 (Vol. 1, Docs. 34 and 35).

[6]In her family history, Alice Rainich Nichols claims that her father was imprisoned because he was teaching relativity. She also describes how her father escaped the prison with the help of his wife and a prison guard who had been a former student of her father. For details, see *Rainich 2013.*

[7]See also *Einstein 1922r* (Vol. 13, Doc. 387, note 2) for an argument as to why the two sets of field equations are indeed not equivalent. The validity of the 1919 trace-free field equations implies the validity of the 1917 field equations with cosmological constant, but not vice versa. However, this did not make a difference for the argument Einstein was making in that paper.

[8]Einstein had already played with this possibility in a draft but then reverted to a purely affine approach in the published version *Einstein 1923e* (Vol. 13, Doc. 425). See Doc. 17, note 3, for details.

[9]For a detailed analysis of how this change in Einstein's thinking came about as a result of the correspondence with Rainich and how it fed into the writing of the paper by Einstein and Grommer, see *Lehmkuhl 2017b.*

[10]See *Lehmkuhl 2017a* for further analysis of Einstein and Grommer's reasoning, especially the question of whether they needed to commit to the existence of singularities.

[11]Einstein would continue to work on this way of approaching the so-called problem of motion in general relativity; his most influential paper on the topic is *Einstein, Infeld, and Hoffmann 1938.* Like Einstein's paper with Grommer, Einstein, Infeld, and Hoffmann aim to derive geodesic motion from the vacuum field equations. The other major approach, which has been widely adopted despite Einstein and Grommer having considered it and opted against it, is to start from the full Einstein equations and derive the geodesic motion of matter from the implied conservation condition of the energy-momentum tensor. For a review of the early work on this problem see *Havas 1989;* for reviews of recent developments see *Asada et al. 2011* and *Puetzfeld et al. 2015.*

[12]This charge has been echoed by *Havas 1989.*

[13]*Yang 1985* argued that Weyl, albeit possibly unwittingly, thus converted Einstein's objection into a prediction, namely, that two electrons which take different paths through an electromagnetic field would end up acquiring different phases in their spin. This is similar to the well-known Aharonov-Bohm effect (see *Aharonov and Bohm 1959*).

[14]See also Teodor Schlomka to Einstein, 2 December 1932 [21 524].

[15]See Schlomka to Einstein, 2 December 1932 [21 524]; Schlomka to Einstein, 30 July 1927 [21 516]; and Einstein to Schlomka, 3 September 1927 [21 518].

[16]See *Lehmkuhl 2014* for an analysis of Einstein's brand of rejecting geometrization as a major message of general relativity; and *Giovanelli 2016* for an analysis of the correspondence between Einstein and Reichenbach on the matter.

[17]See Einstein to Eduard Einstein, 25 June 1923 (Vol. 14, Doc. 68).

[18]Frieda was nine years older than Hans Albert, whereas Mileva was only four years older than Einstein.

[19]The vehement opposition of Einstein's parents, in particular of his mother, to his plans to marry Mileva, was based on her allegedly not descending from a "respectable family" and on her being older than Einstein (Vol. 1, Introduction, p. xxxvii, and Einstein to Mileva Marić, 29? July 1900 [Vol. 1, Doc. 68]).

[20]In 1917, Einstein had expressed remorse and self-reproach for having fathered children with Mileva, "a physically and morally inferior person." Yet, at the same time, he admitted that his own

family lacked a high-quality genetic pedigree (Einstein to Heinrich Zangger, 16 February 1917 [Vol. 8, 299a, in Vol. 10]).

[21]See *Ettema and Mutel 2014*, p. 59.

[22]On Eduard during the period of this volume, see *Rübel 1986*, pp. 23–56.

[23]See Vol. 14, Introduction, pp. lii–liv.

[24]On Einstein's affair with Betty Neumann, see Vol. 14, Introduction, pp. lv–lvi.

[25]See Doc. 469, note 5.

[26]See Vol. 7, "An Interview with Professor Albert Einstein" (Appendix D), pp. 620 and 624.

[27]See *Cleveland Press*, 25 May 1921; Miller to T. C. Mendelhaft, 2 June, American Institute of Physics Archive [70 984].

[28]Later experts who reanalyzed his data also concluded that Miller's results are consistent with variations caused by temperature changes (see *Shankland et al. 1955*).

[29]See *Illy 2006*, p. 303, for a drawing of that day by Einstein and Miller's notes.

[30]On Herschel, see also *Hoskin 1980*.

[31]On the Locarno Treaties, see *Vincent 1997*, pp. 286–287.

[32]On Germany's entrance into the League of Nations, see *Winkler 1993*, pp. 308 and 315, and *Wintzer 2006*, pp. 501–504.

[33]On the government crisis, see *Winkler 1993*, pp. 308–310.

[34]On the issue of the former princes' property, see *Winkler 1993*, pp. 312–315.

[35]On the economic crisis, see *Hardach 1980*, p. 31, *Braun, H. 1990*, p. 47, and *Winkler 1993*, p. 311.

[36]See Vol. 14, Introduction, pp. xlv–xlvi.

[37]On the judicial system during the Weimar Republic and its prosecution of left-wing groups, see *Hannover and Hannover-Drück 1966*, pp. 192–250, *Olenhusen 1971*, and *Evans 2003*, pp. 134–138.

[38]For the planning of these events, see Vol. 14, Introduction, pp. lii–liv

[39]On the issue of capital punishment in the Weimar Republic, see *Evans 1996*, pp. 499–561, and *Hammel 2010*, pp. 62–63.

[40]This use of biological terminology that evokes comparisons with eugenic thinking is perhaps less surprising given the context of some of Einstein's previous statements on the value of life. In 1917, Einstein believed that his own offspring was genetically "inferior" (see note 20), when he even raised the possibility of imitating "the methods of the Spartans" to deal with Eduard's alleged inferiority. A year later, he differentiated between "valuable people," i.e., those with superior intellect, and those who were less valuable, i.e., "unimaginative average [people]," and who were therefore more expendable in war (see Einstein to Otto Heinrich Warburg, 23 March 1918 [Vol. 8, Doc. 491]). A further instance of similar judgments related to his close friend Paul Ehrenfest's youngest son, Wassily, having been diagnosed with Down syndrome. Einstein approved of the plan "to hand the child over to impersonal care," and added that "valuable people should not be sacrificed to hopeless causes" (see Einstein to Paul Ehrenfest, on or after 22 August 1922 [Vol. 13, Doc. 329]). For more on the limits of Einstein's humanism, see *Rosenkranz 2011*, pp. 42–44, 249–250, and 266–267.

[41]See, e.g., Einstein's willingness to assist Austrian physicist and social democrat Friedrich Adler in avoiding the death penalty for assassinating the Austrian prime minister Count Karl von Stürgkh in 1916, and his support for an appeal against the possible execution of the German communist leader Eugen Leviné in 1919 (see Einstein to Friedrich Adler, 13 April 1917 [Vol. 8, Doc. 324], and Einstein to Munich Military Tribunal, 19 May 1919 [Vol. 9, Doc. 44]).

[42]On Einstein's elitism, see *Rosenkranz 2011*, pp. 43–44 and 184.

[43]See *Zimmermann 1997*, pp. 40–43.

[44]In contrast to Weizmann and the Labor Zionists, the right-wing and revisionist Zionists opposed public entrepreneurship and the centrality of the Labor movement in agricultural settlement enterprises. Instead, they aimed to represent middle-class interests and advocated private economic ventures, especially in urban areas. Furthermore, the revisionists objected to the concept that the Zionist Organisation should be the central body for Jewish settlement in Palestine. They believed the British mandate authorities should provide the framework and infrastructure for a Jewish "national home" by establishing a "colonization regime." On the political opposition to Weizmann's policies, see *Lavsky 1996*, pp. 115–116 and 122–125.

[45]See *Freundlich 1977*, p. 396, note 3.

[46]See *Lavsky 1996*, pp. 116–120 and 162–176, and *Zimmermann 1997*, pp. 33–34.

[47]On Einstein's previous involvement with the Hebrew University, see Vol. 14, Introduction, pp. lxxii–lxxiii and lxxvii–lxxviii. On the celebrations in Buenos Aires, see Vol. 14, Introduction, p. lxx–lxii, and Appendix H).

[48]See "Protokoll der Zweiten Sitzung des Kuratoriums der Universitaet Jersualem, abgehalten am 23. Sept. 1925 in Muenchen" (IL-JeCZA, L12/83/1/1; 89 854).

[49]In British universities, the chancellor is the representative head, and the vice-chancellor is the one who holds the actual political power and is in effect head of the executive, to whom everybody else reports.

[50]On the conflict between Einstein and Magnes on the governance of the Hebrew University, see *Parzen 1974*, pp. 5–7, *Goren 1996*, pp. 213–218, *Rosenkranz 2011*, pp. 189–199.

[51]For more details, see Doc. 142, note 1. The first version of the minutes is extant only in German (see "Protokoll der Zweiten Sitzung des Kuratoriums der Universitaet Jerusalem, abgehalten am 23. Sept. 1925 in Muenchen" [IL-JeCZA, L12/83/1/1]). The second version is extant in a draft in English and as a fragment in Hebrew (see "Draft. Minutes of the second Meeting of the Board of Governors of the Hebrew University Held at Munich in the house of Dr. Eli Strauss September 23rd and 24th, 1925" [OCAJA, Felix M. Warburg Papers, MS-457, box 220, folder 6], and "Minutes of the Second Meeting of the Board of Governors of the Hebrew University held in Munich at the home of Dr. Eli Straus, 23 and 24 September 1925" (in Hebrew) [IL-JeHUCA]). The editors of *Freundlich 1977* claimed that the first version of the minutes had "not been traced" (see *Freundlich 1977*, p. 419).

[52]On their previous contacts, see Vol. 12, Introduction, pp. xxix–xxx, and Vol. 13, Introduction, p. lxv.

[53]On the reestablishment of the Deutsches Komitee "Pro Palästina," see *Walk 1976*.

[54]Decades later, philosophers began to discuss the question of whether one should deduce and embrace ontological consequences from the fact that the wave function is defined on configuration space rather than on spacetime. Those who answer this question affirmatively often call themselves "wave function realists." For sources see *Ney and Albert 2013*.

[55]For the distinction between different kinds of hidden variable theories, see *Barrett, J. 1999*; for detailed analysis of Einstein's manuscript (Doc. 516), see *Belousek 1997* and *Holland 2005*. Belousek and Holland disagree on what the hidden variables in Einstein's theory are: Belousek argues that it is the principal directions $\{\lambda_{(a)}\}$ that Einstein introduces, whereas Holland argues that these are functions of $\psi$ and thus not hidden variables. Instead, he takes the positions of the particles as the hidden variables of the theory.

[56]*Dannen 1997*. For their relationship, see *Lanouette 1992*, ch. 6.

[57]For a suggestion as to how this pump might have worked, see *Graff 2004*, p. 227.

[58]S77558 IVb/12a was patented as DE525833.

[59]See Vol. 14, Introduction, p. lxxii.

[60]For Einstein declining Millikan's previous invitations to Caltech, see Einstein to Robert A. Millikan, 3 November 1924 (Vol. 14, Doc. 360) and Doc. 20.